\documentclass[useAMS,usenatbib]{mn2e}

\usepackage{epsfig,amsmath,fleqn,amssymb,color}

\title[Warp and eccentricity propagation]{Warp and eccentricity propagation in discs around black holes}
\author[B\'arbara T. Ferreira and Gordon I. Ogilvie]{B\'arbara T. Ferreira$^{1}$\thanks{E-mail:
B.T.Ferreira@damtp.cam.ac.uk} and Gordon I. Ogilvie$^{1}$\\
$^{1}$Department of Applied Mathematics and Theoretical Physics, University of Cambridge, Wilberforce Road, Cambridge CB3 0WA}
\begin{document}

\date{Accepted 2008 . Received 2008 ; in original form 2008 September 13}

\pagerange{\pageref{firstpage}--\pageref{lastpage}} \pubyear{2008}

\maketitle

\label{firstpage}

\begin{abstract}
We consider the inward propagation of warping and eccentric disturbances in discs around black holes under a wide variety of conditions. In our calculations we use secular theories of warped and eccentric discs and assume the deformations to be stationary and propagating in a disc model similar to regions (a) and (b) of Shakura and Sunyaev discs. We find that the propagation of deformations to the innermost regions of the disc is facilitated for low viscous damping and high accretion rate. We relate our results to the possible excitation of trapped inertial modes, and to the observations of high-frequency quasi-periodic oscillations (QPOs) in black hole systems in the very high spectral state.
\end{abstract}


\section{Introduction}

\subsection{Background}

In the classical theory of accretion discs the gas is assumed to rotate around a massive central object following coplanar, circular and approximately Keplerian orbits. Although this is the simplest solution for the motion of fluid elements placed in a Newtonian potential well, external forces can give rise to non-planar discs, composed of non-circular rings. In fact, the general solution allows for discs to be twisted or tilted, with their orbits presenting a smoothly varying eccentricity and/or inclination. The literature offers strong observational and theoretical evidence \citep[][and references therein]{bardeenpetterson1975,ogilvie2000,ogilvie2001,globalsimulations} to believe that under various conditions discs can become either warped or eccentric. In particular, the precession of deformed discs has been used to explain several phenomena in X-ray binaries \citep[e.g.][]{gerendboynton1976,whitehurst1988,sv1998,lai1999}.
  
Discs around black holes are warped if the spin axis of the central object and the angular momentum of the accreting matter are misaligned. The fluid elements in tilted orbits will be subject to Lense-Thirring (gravitomagnetic) precession which tends to twist and warp the disc \citep{bardeenpetterson1975}. The way a tilted ring communicates its inclination with respect to the equator of the black hole to its neighbouring rings depends on a balance between hydrodynamic and gravitomagnetic torques. Roughly speaking the warp propagation is diffusive if $\alpha>H/r$, and is wave-like if $\alpha<H/r$, where $\alpha$ is the \cite{ss73} viscosity parameter and $H/r$ is the angular semi-thickness of the disc. In a low-viscosity disc around a Kerr black hole, the warp has an oscillatory radial structure, as shown by \cite{ivanovillarionov1997} and discussed by \cite{lubowetal2002}. In several X-ray binaries, jets (which are thought to be aligned with the rotation axis of the black hole) are observed to be misaligned with respect to the binary rotation axis \citep[e.g. GRO J1655-40,][]{maccarone2002}. This is strong evidence for a warped disc around the compact object in the binary \citep{mtp2008}. Warping might also be induced by radiation pressure forces \citep{pringle1996}. Despite being more likely for discs around neutron star primaries \citep{ogilviedubus2001}, radiation-driven warping might also occur around black holes provided the disc extends to large enough radius, as in the case of GRS 1915+105 \citep{rauetal2003}.

Eccentric accretion discs are believed to exist in cataclysmic binaries of mass ratio $q\la 0.3$ \citep[e.g.][]{pattersonetal2005}. Eccentricity can result from an instability involving the orbiting gas and the tidal potential of the companion star. Discs may become eccentric if they are large enough to extend to the 3:1 resonance, the radius in the disc where its angular velocity equals three times that of the binary \citep{lubow1991a,lubow1991b}. The strongest observational evidence for eccentric discs is the detection, in the light curves of accreting binary systems, of long-period modulations known as superhumps, which can be explained by the action of tidal stresses on a precessing eccentric disc \citep{whitehurst1988}. Although this phenomenon is usually associated with cataclysmic variable stars, superhumps have been detected in an increasing number of black hole binaries \citep{donocharles1996,haswelletal2001,uemura2002,neiletal2007,zurita2008}. In fact, black hole binaries are likely to have mass ratios $q\la0.3$ and therefore to have eccentric discs during at least some stages of their outbursts. \cite{haswelletal2001} noted that the mechanism responsible for superhump luminosity variations differs in X-ray binaries and cataclysmic variables, but the underlying dynamics is the same.

Our goal is to describe the propagation of global eccentric and warped disturbances, under a variety of conditions, from the outer parts of the disc where they originate to the inner parts. In the outer parts of the disc, the motion is Keplerian and stationary or slowly precessing global deformations are supported. Since the Lense-Thirring precession frequency increases with decreasing $r$, tilted orbits at different radius will tend to precess at different rates, which tends to twist the disc. Hydrodynamic stresses counteract this effect, resulting in wavelike warp propagation from the outer disc to the inner region. Similarly, for eccentric discs a transition has to be made between the Keplerian region where eccentric instabilities are driven, and the inner region where general relativistic effects dominate the precession of elliptical orbits. It is this connection between different regions of the disc, with different characteristics, that we attempt to describe.

\subsection{Deformations as global modes}

Warped or eccentric disturbances of small amplitude can be thought of as linear perturbations of a circular and coplanar disc. In the simplest case of a vertically isothermal disc with unit ratio of specific heats, wave modes with a dependence in time and azimuth of the form $\exp{(\textrm{i}m\phi-\textrm{i}\omega t)}$, and local radial wavenumber $k$ obey \citep{okazakietal1987}

\begin{equation}
k^2=\frac{(\hat{\omega}^2-\kappa^2)(\hat{\omega}^2-n\Omega_z^2)}{\hat{\omega}^2c_\mathrm{s}^2},
\label{disprelation}
\end{equation}
where $\hat{\omega}=\omega-m\Omega$ is the Doppler-shifted wave frequency, integers $m$ and $n\geq0$ are the azimuthal and vertical mode numbers, respectively, and $c_\textrm{s}=\Omega_z H$ is the constant sound speed in the disc, with $H$ being the local vertical scaleheight. The symbols $\Omega$, $\kappa$ and $\Omega_z$ represent the angular velocity, epicyclic frequency and vertical frequency, respectively. Relativistic effects are included in the problem by using relativistic expressions for these characteristic frequencies. In the Kerr metric, valid for a rotating black hole, they read \citep{kato1990} 

\begin{equation}
\Omega=(r^{3/2}+a)^{-1},
\label{relomega}
\end{equation}
\begin{equation}
\kappa=\Omega\sqrt{1-\frac{6}{r}+\frac{8a}{r^{3/2}}-\frac{3a^2}{r^2}},
\label{relkappa}
\end{equation}
\begin{equation}
\Omega_z=\Omega\sqrt{1-\frac{4a}{r^{3/2}}+\frac{3a^2}{r^2}},
\label{relomegaz}
\end{equation}
where $a$ is the dimensionless spin parameter of the black hole
($-1<a<1$), $r$ is the cylindrical radius in units of $GM/c^2$, and the frequencies are in units of $c^3/GM$, where $c$ is the speed of light, $G$ the gravitational constant and $M$ the mass of the black hole; in these units velocities are in units of $c$. The dispersion relation is local, i.e., it is different at each radius since the characteristic frequencies depend on $r$, and is valid only where the wavelength $\lambda=2\pi/k\ll r$.


\begin{figure}
\begin{center}
\includegraphics[width=\linewidth]{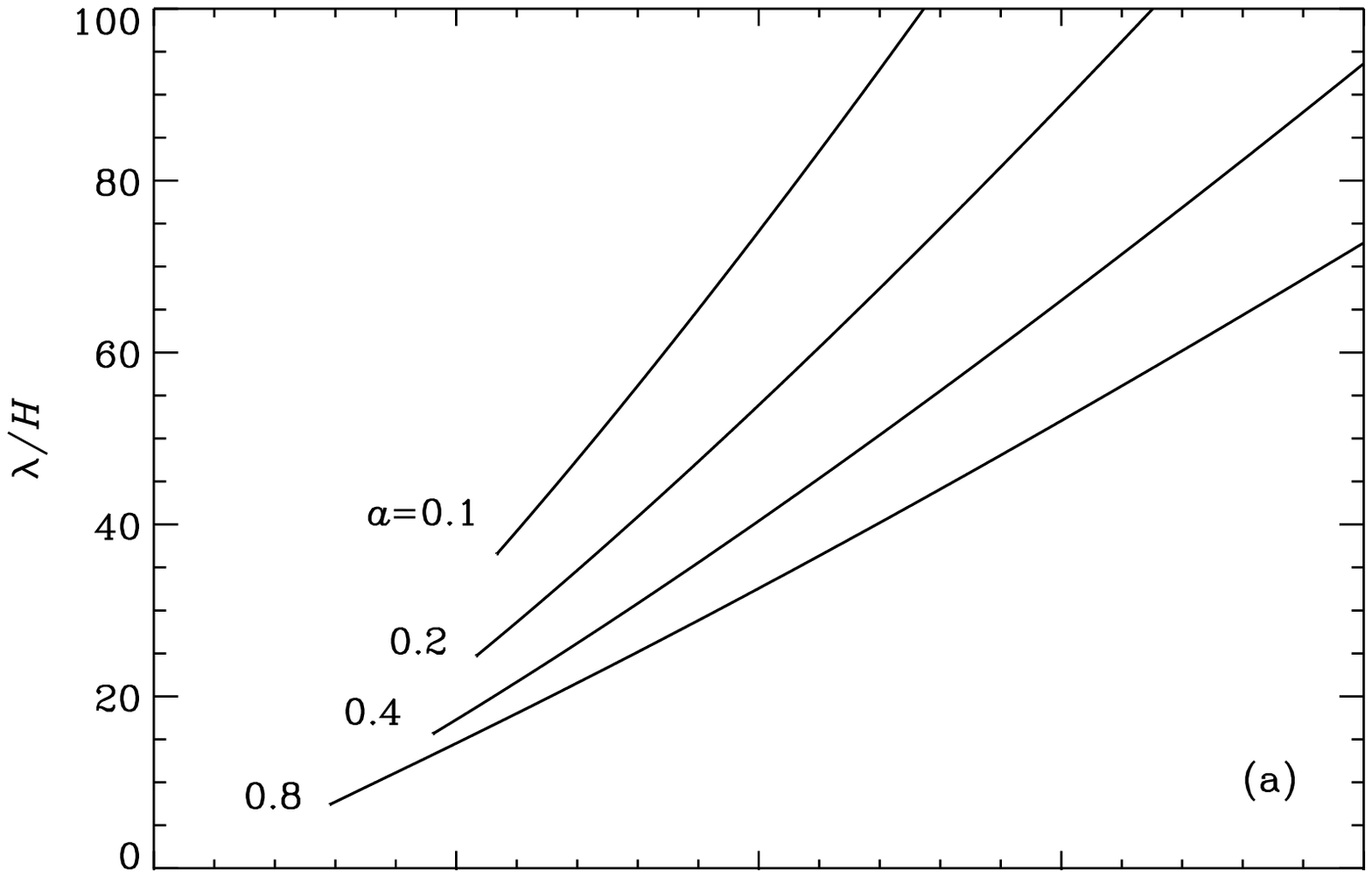} \\
\vspace{-5mm}
\includegraphics[width=\linewidth]{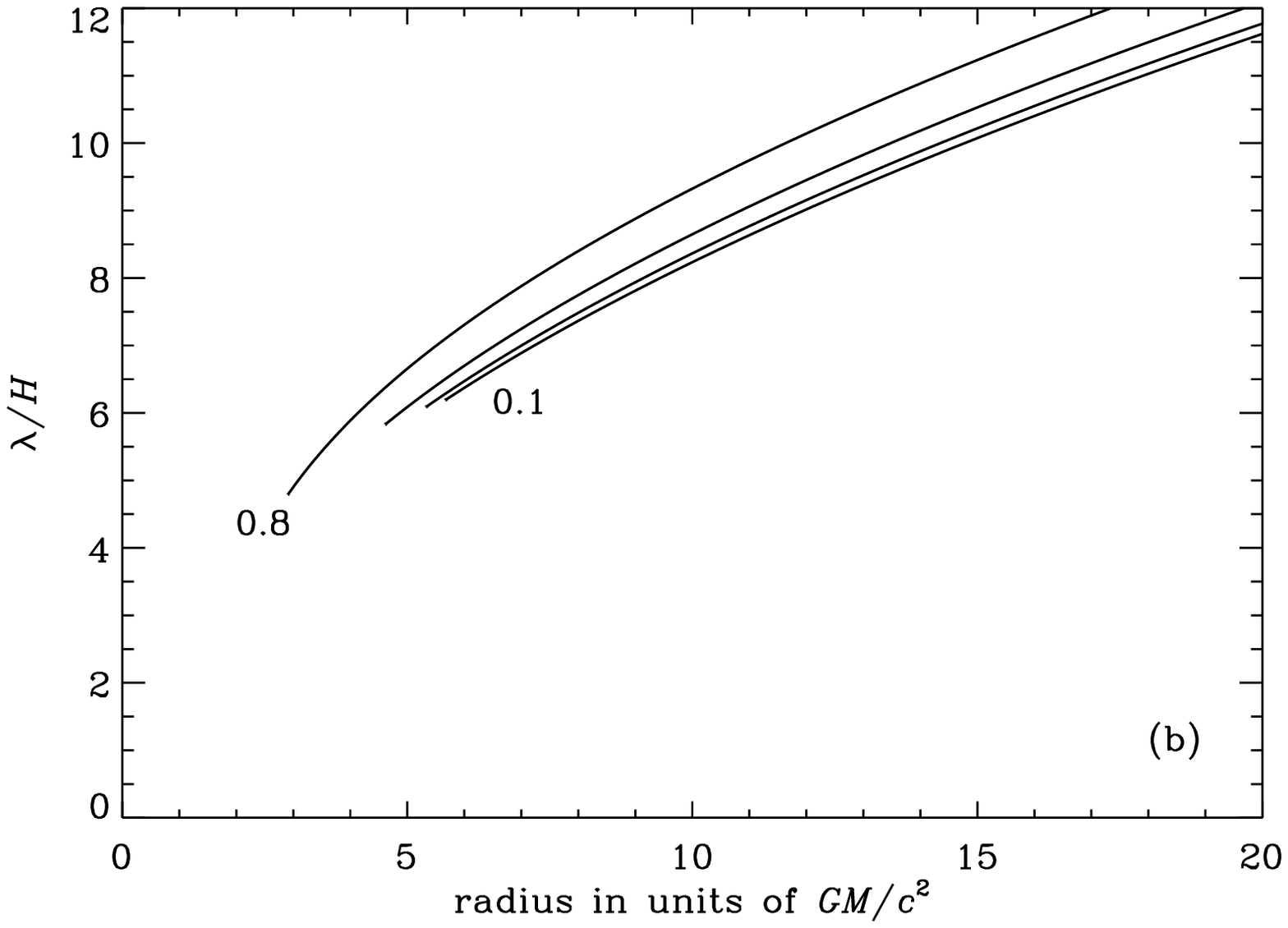}
\end{center}
\caption{Local wavelength of (a) $(\omega,m,n)=(0,1,1)$ and (b) $(0,1,0)$ modes (stationary warp and eccentricity, respectively) in units of $H$ for several values of the black hole spin. The inner edge of the disc is taken to be the marginally stable orbit, where $\kappa=0$.}
\label{lambdah}
\end{figure}

Viewed as linear perturbations of a standard disc, warp and eccentricity correspond to low-frequency waves with azimuthal number $m=1$. As showed by \cite{kato1983,kato1989} these modes are global, and have relatively long wavelengths when compared to the other modes described by the local dispersion relation, being the most likely to exist in a turbulent disc. More specifically, a $(m,n)=(1,1)$ mode can be identified with a warp as its vertical displacement is independent of the vertical coordinate $z$ and varies with $\phi$ at each radius, giving rise to a tilting (movement with respect to the disc plane). A $(m,n)=(1,0)$ mode corresponds to eccentric disc since its radial displacement is independent of $z$ and varies with azimuthal angle, originating an elliptic orbit (movement within the disc plane). The $m=1$ modes are global in the sense that only they can have a wavelength much larger than the semi-thickness of the disc over a wide range of radius. This can be confirmed in Fig. \ref{lambdah} where we show the variation with $r$ of the local wavelength in units of $H$, for several values of $a$, for both the warp and eccentricity.

In \cite{ferreiraogilvie2008} we worked with a simplified set of hydrodynamic equations in order to deal with the nonlinear couplings of different wave modes. Within this framework we obtained an equation for the warp tilt $W$ at each radius \citep[similar to the one given by][]{papaloizoulin1995} assuming the warp to be a zero-frequency mode propagating in a strictly isothermal, relativistic disc. In addition to being valid only where $c_\textrm{s}$ is constant and where viscous effects are negligible, the equation derived does not hold for large radius, where the wavelength of the warp becomes comparable to $r$, since some terms were omitted in our local approximation. A similar method was used to derive an equation for the eccentricity at each radius $E$ valid only for an inviscid and strictly isothermal disc.

Here we consider the more general description of a stationary, wave-like warp and eccentricity given by \cite{lubowetal2002} and \cite{goodchildogilvie2006}, respectively. These theories are less general than the ones formulated by \cite{ogilvie2000,ogilvie2001} since they consider deformations to have small amplitude, and not all viscous or turbulent effects are included. Also, the \cite{goodchildogilvie2006} calculation uses a highly simplified, 2D disc model. However, the secular theories of \cite{lubowetal2002} and \cite{goodchildogilvie2006} are appropriate for thin discs of any given structure, and allow for viscous (turbulent) damping of deformations.

\subsection{Black hole states and high-frequency QPOs}

Observations indicate that black hole systems can be found in different spectral states, going from the quiescent state to low, intermediate, high and very high states as the mass accretion rate increases \citep{bhbbook}. The very high state is the one where high-frequency QPOs, i.e., modulations seen in the spectra of black hole systems that are not strictly periodic and have frequencies $\ga 100$ Hz, are almost exclusively detected \footnote{In the X-ray binary XTEJ1550-564, a HFQPO was detected at a photon count rate below the VHS \citep[see][and references therein]{klisbook}.}. This is the dominating state close to the Eddington limit. 

The accretion flow has different characteristics in different black hole states. If the disc properties are such that the warp and eccentricity can propagate to the inner region, the interaction between relativistic effects and global deformations can give rise to interesting phenomena. In particular, the warp and eccentricity can have a fundamental role in exciting trapped inertial modes, which may explain high-frequency QPOs \citep{nowaketal1997}, as previously suggested by \cite{katowarp2004,kato2008} and \cite{ferreiraogilvie2008}. 

In this paper we show that the inward propagation of warp and eccentricity is facilitated for high accretion rate. For fixed viscosity, global deformations reach the inner region of the disc with a modest amplitude when the accretion rate is high, and can take part in the excitation mechanism for inertial modes trapped in this region. If high-frequency QPOs can be identified with these trapped inertial waves, we can argue that they are predominantly detected in the very high state, where the accretion rate is close to Eddington, because only in this case global deformations can reach the inner region with non-negligible amplitude and excite the trapped modes.

\subsection{Plan of the paper} 

In section 2, we introduce the disc model in which we study the propagation of global deformations. We consider a disc with a polytropic structure in the vertical direction, constant opacity, and with gas and radiation contributing to the total pressure. In section 3 we describe the propagation of global disturbances, by solving equations for the warp tilt and eccentricity at each radius. To mimic several possible conditions in a disc with a fixed viscosity $\alpha$ around a black hole of a given mass and spin, we vary the accretion rate and the damping of the warp and eccentricity. The results obtained are discussed in section 4. Conclusions are presented in section 5.

\section{Disc model}

\subsection{Viscosity prescription}

In the famous \cite{ss73} model, the disc is assumed to be geometrically thin and optically thick. The gas in the disc is heated by viscous dissipation and cooled by radiative diffusion through the vertical boundaries. Angular momentum is transferred outwards at a rate which is taken to be proportional to the height-integrated total pressure, with the efficiency of the process being parametrised by a dimensionless quantity $\alpha$, independently of the nature of the stress. The disc is taken to be composed of three parts: in the innermost regions (a) and (b), electron scattering is the main contribution to the opacity (which is then given by the constant, Thomson value, $\kappa_{\textrm{T}}$), while free-free absorption is dominant in the outer region (c). In (a) the pressure is determined by the radiation pressure ($p\propto T^4$), while gas pressure ($p\propto \rho T$) dominates in (b) and (c). The radiation-dominated region exists only if the accretion rate is larger than a few percent of the Eddington rate, being important in accretion discs around black holes.

Despite neglecting to treat the vertical structure of the disc, not considering other contributions to the cooling and heating rates of the gas, and despite the simple parametrization for the angular momentum transfer, the Shakura \& Sunyaev $\alpha$-model is still the standard reference for the radial structure of accretion discs, if gas pressure is dominant. Region (a) has always been more problematic. In fact, the assumption that stress scales with the total pressure in the radiation pressure dominated regime leads to viscous \citep{le74} and thermal \citep{ss76} instabilities. These can be avoided by simply assuming that the stresses scale with the gas pressure \citep{sc81}, or with a combination of both gas and radiation pressure \citep[e.g.][]{tl1984,mf2002}, recent simulations by Hirose et al. (ApJ, submitted) suggest that turbulent stresses may indeed be correlated with the total pressure. However, fluctuations in stress \citep[now known to be magnetic in nature,][]{mri} are the ones driving pressure fluctuations, contrary to the usual assumption that changes in pressure give rise to changes in stress; this seems to avoid thermal instability (which has a faster growth rate than viscous instability). With these results in mind, we consider the $\alpha$ viscosity prescription in the form,

\begin{equation}
\mu=\alpha\frac{p}{\Omega},
\label{avp}
\end{equation}
so that the viscosity $\mu$ and total pressure $p$ correlate, as in the original Shakura \& Sunyaev model. Here we consider $\alpha$ to be a function of radius only; in this section $\Omega=\Omega_K=\sqrt{GM/R^3}$ is the Keplerian angular velocity, and $R=rGM/c^2$ is the dimensional radius.

\subsection{Vertical structure}

Independently of the viscosity prescription, the radiation-pressure dominated region might suffer from another type of instability. In the Shakura \& Sunyaev model, the dissipation rate per unit volume is independent of $z$. If the dissipation rate per unit mass is also vertically constant, then the density is independent of $z$, vanishing abruptly at the vertical boundaries of the disc, $z=z_0(R)$ \citep{ss73}. 
A disc with these properties is subject to convective instability \citep{bb77}.

Simulations of radiation-dominated discs by \cite{turner2004} indicate that neither is the dissipation per unit mass constant nor is the density independent of $z$. In fact, the computed density profile resembles more that of a polytropic model with index $s=3$ \citep[see also][]{agoletal2001}. Using the polytropic law, $p\propto \rho^{4/3}$, in the hydrostratic equilibrium equation 

\begin{equation}
\frac{\partial p}{\partial z}=-\Omega_z^2\rho z \quad (\textrm{Keplerian disc: $\Omega_z=\Omega=\sqrt{GM/R^3}$}),
\label{he}
\end{equation}
it is straightforward to get

\begin{equation}
\rho=\rho_0(R) \left[1-\left(\frac{z}{z_0}\right)^2\right]^3 \quad (-z_0<z<z_0).
\label{rho}
\end{equation}
This result was used by \cite{bb77}, who argued that convection in the radiation-pressure dominated region would establish a vertically isentropic structure ($T^3/\rho=$ constant in $z$).

The $s=3$ polytropic structure has the convenient property of allowing for a ratio of radiation to gas pressure,

\begin{equation}
\beta=\frac{p_\textrm{r}}{p_\textrm{g}}=\frac{4\sigma\mu_\textrm{m} m_{\textrm{p}}}{3ck_B}\frac{T^3}{\rho}, 
\end{equation}
independent of $z$. Using (\ref{rho}) in the hydrostratic balance equation we get $p=p_0(R)\left[1-\left(z/z_0\right)^2\right]^4$, where

\begin{equation}
p_0=\frac{\Omega_z^2z_0^2}{8}\rho_0=\frac{\Omega^2z_0^2}{8}\rho_0.
\label{he2}
\end{equation}
This is consistent with the equation of state

\begin{equation}
p=p_{\textrm{g}}+p_{\textrm{r}}=\frac{k_B}{\mu_\textrm{m} m_{\textrm{p}}}\rho T+\frac{4\sigma}{3c}T^4,
\end{equation}
provided $T=T_0(R)\left[1-\left(z/z_0\right)^2\right]$, and independently of the variation of $\beta$ with radius, i.e., the polytropic vertical structure with $s=3$ can be used to model not only the radiation pressure dominated regime but also the gas pressure dominated region. In fact, a full treatment of the vertical structure of a gas pressure dominated disc in hydrostatic (\ref{he}) and radiative equilibrium \citep[e.g.][]{accretionpower}:

\begin{equation}
\frac{\partial F}{\partial z}=\mu\left(R\Omega'\right)^2 \quad \textrm{(energy balance)},
\label{eb}
\end{equation}
\begin{equation}
F=F_{\mathrm{rad}}=-\frac{16\sigma T^3}{3\kappa_\textrm{T}\rho}\frac{\partial T}{\partial z} \quad \textrm{(radiative diffusion law)},
\label{rdl}
\end{equation}
indicates that the resulting profiles are very similar (except perhaps $F(z)$ for large $z$) to those obtained for a polytropic model with $s\approx 2.7$ (Ogilvie 2001, unpublished lecture notes). This result, and the simulations in radiation-pressure dominated regions, indicate that assuming a polytropic vertical structure with $s=3$ throughout the radial extent of the disc is a good approximation, provided the opacity remains constant. This is true of the innermost regions ((a) and (b) in the Shakura \& Sunyaev model) of the disc. Since we are most interested in the propagation of global deformations in these regions, we take the opacity to be constant throughout the disc, i.e., region (c) is neglected in our model. In fact, the physics of region (c) is more complicated than assumed by \cite{ss73}, because of the presence of different sources of opacity and the importance of irradiation and partial ionisation.

\subsection{Radial structure}

The assumptions of hydrostatic equilibrium and vertical isentropy (or equivalently, $s=3$ polytropic structure in $z$), together with the equation of state, are enough to determine the vertical structure of the disc, i.e., determine the variation of $\rho$, $p$ and $T$ with $z$, without the requirement of radiative balance \citep[see discussion in section 3.1 of][]{agoletal2001}. However, to fully determine the radial structure: $\rho_0(R)$, $p_0(R)$, $T_0(R)$ and $z_0(R)$, two more equations are needed since the polytropic relation is not valid in the radial direction.

Since matter is being accreted in the disc, there are radial drift motions which transport mass and angular momentum. In steady thin discs the conservation of both mass and angular momentum can be express in the form \citep[e.g.][]{pringle1981},
\begin{equation}
\bar{\nu}\Sigma=\frac{\dot{M}}{3\pi}f, \quad \textrm{with}\quad f=1-\sqrt{\frac{R_\textrm{in}}{R}},
\label{cam}
\end{equation}
where $\dot{M}$ is the mass accretion rate, $R_\textrm{in}$ is the radius of the inner edge of the disc taken to be the marginally stable orbit, and the (density-weighted) mean kinematic viscosity $\bar{\nu}(R)$ is defined by

\begin{equation}
\bar{\nu}\Sigma(R)=\int^{z_0}_{-z_0}\mu(R,z) \textrm{d}z=\alpha\frac{P}{\Omega},
\end{equation}
where

\begin{equation}
\Sigma(R)=\int^{z_0}_{-z_0}\rho(R,z)\textrm{d}z=\frac{32}{35}\rho_0z_0,
\end{equation}
\begin{equation}
P(R)=\int^{z_0}_{-z_0}p(R,z) \textrm{d}z=\frac{256}{315}p_0z_0
\end{equation}
are the surface density and vertically integrated pressure, respectively.

The final equation comes from energy considerations. It is traditional to assume that the disc is in radiative equilibrium. However, convection, turbulence or other motions can also contribute to the transport of energy to the disc surface. In fact, if the disc has a polytropic vertical structure with $s=3$, the radiative diffusion law gives us $F_{\textrm{rad}}$ proportional to $z$. If $F=F_{\textrm{rad}}$, i.e., if radiative diffusion carries the entire energy flux, then (\ref{eb}) gives us a dissipation rate per unit volume independent of $z$, which is neither consistent with the $\alpha$ viscosity prescription assumed here (\ref{avp}) since $p=p_0(R)\left[1-\left(z/z_0\right)^2\right]^4$, nor is in agreement with simulations of radiation-pressure dominated discs \citep{turner2004}. Therefore, here we use the energy balance equation in the form

\begin{equation}
\int_{0}^{z_0}\frac{\partial}{\partial z}\left(F_{\textrm{rad}}+F_{\textrm{extra}}\right)\textrm{d}z=\int_{0}^{z_0}\mu\left(R\Omega'\right)^2\textrm{d}z,
\end{equation}
where the energy transport by other motions rather than by radiative diffusion is encompassed in $F_{\textrm{extra}}$. Since radiation is supposed to carry the entire heat flux at the photosphere $z=z_0$, we assume that the extra term integrates to zero so that the energy balance can be written as,

\begin{equation}
F_{\textrm{rad}}(R,z_0)=\frac{1}{2}(R\Omega')^2\left(\alpha\frac{P}{\Omega}\right),
\end{equation} 
since $F_{\textrm{rad}}(R,z=0)=0$ by symmetry.



Here we take $\mu_\textrm{m}=0.615$, $\kappa_T=0.33\textrm{ cm}^2\textrm{g}^{-1}$, and use non-dimensional parameters to represent the dependence in mass, accretion rate and radius. The radial structure of the disc can be represented in terms of $\Sigma(R)$, $P(R)$, $\beta(R)$ and $H(R)$, the (density-weighted) scale-height of the disc defined by:

\begin{equation}
H^2=\frac{\int_{-z_0}^{z_0}\rho z^2\textrm{d}z}{\int_{-z_0}^{z_0}\rho\textrm{d}z}=\frac{z_0^2}{9}.
\label{dt}
\end{equation}
The equations read:
\begin{equation}
\Sigma=2.5\times10^5\, \dot{m}^{3/5}\alpha^{-4/5}m^{1/5}f^{3/5}r^{-3/5} (1+\beta)^{-4/5},
\label{rsc}
\end{equation} 
\begin{equation}
P=3.0\times10^{22}\,\dot{m}\,\alpha^{-1}fr^{-3/2},
\label{rsd}
\end{equation} 
\begin{equation}
\beta (1+\beta)^{-3/5}=3.6\times10^2\, \dot{m}^{4/5}\alpha^{1/10}m^{1/10}f^{4/5}r^{-21/20},
\label{rsa}
\end{equation}
\begin{equation}
H=1.7\times10^3\,  \dot{m}^{1/5}\alpha^{-1/10}m^{9/10}f^{1/5}r^{21/20} (1+\beta)^{2/5},
\label{rsb}
\end{equation} 
where $m$ is the mass in units of $M_\odot$, $\dot{m}$ is the accretion rate in units of the Eddington accretion rate assuming an efficiency of $0.1$, $r$ is in units of the gravitational radius and $f=1-\sqrt{r_\textrm{in}/r}$, where $r_\textrm{in}$ is the dimensionless, spin dependent, radius of the marginally stable orbit; $H$, $\Sigma$ and $P$ are in cgs units. It should be noted that in the limits $\beta\gg1$ ($P_{\textrm{r}}$ dominates) and $\beta\ll1$ ($P_{\textrm{g}}$ dominates) we recover the same dependencies in $m$, $\dot{m}$, $\alpha$, $r$ and $f$ as in regions (a) and (b), respectively, of the Shakura \& Sunyaev model:\newline
\newline
\textbf{(a) $\mathbf{\beta\gg1}$:}
\begin{equation}
\Sigma=1.9\,\alpha^{-1}\dot{m}^{-1}r^{3/2}f^{-1},
\end{equation}
\begin{equation}
\beta =2.4\times10^6\, \dot{m}^{2}\alpha^{1/4}m^{1/4}f^{2}r^{-21/8},
\end{equation}
\begin{equation}
H=6.1\times10^5\dot{m}\,m\,f.
\label{ha}
\end{equation}
\vspace{5pt}
\textbf{(b) $\mathbf{\beta\ll1}$:}
\begin{equation}
\Sigma=2.5\times10^5\,\alpha^{-4/5}\dot{m}^{3/5}m^{1/5}r^{-3/5}f^{3/5},
\label{sdb}
\end{equation}
\begin{equation}
\beta=3.6\times10^2\, \dot{m}^{4/5}\alpha^{1/10}m^{1/10}f^{4/5}r^{-21/20},
\end{equation}
\begin{equation}
H=1.7\times10^3\,\alpha^{-1/10}\dot{m}^{1/5}m^{9/10}r^{21/20}f^{1/5}.
\label{hb}
\end{equation}
The expression for the total pressure in both regions is the same and is given by (\ref{rsd}). The differences in the numerical factors are due to the treatment of the vertical structure, to the small difference in the values used for $\kappa_T$ and $\mu_\textrm{m}$, to the different normalization of radius, and also because here we consider the $r\phi$ component of the stress tensor, $T_{r\phi}=\mu r\Omega'$, to be $-(3/2)\alpha p$, instead of the usual Shakura \& Sunyaev value $-\alpha p$.

\begin{figure*}
\begin{center}
\includegraphics[width=0.48\linewidth]{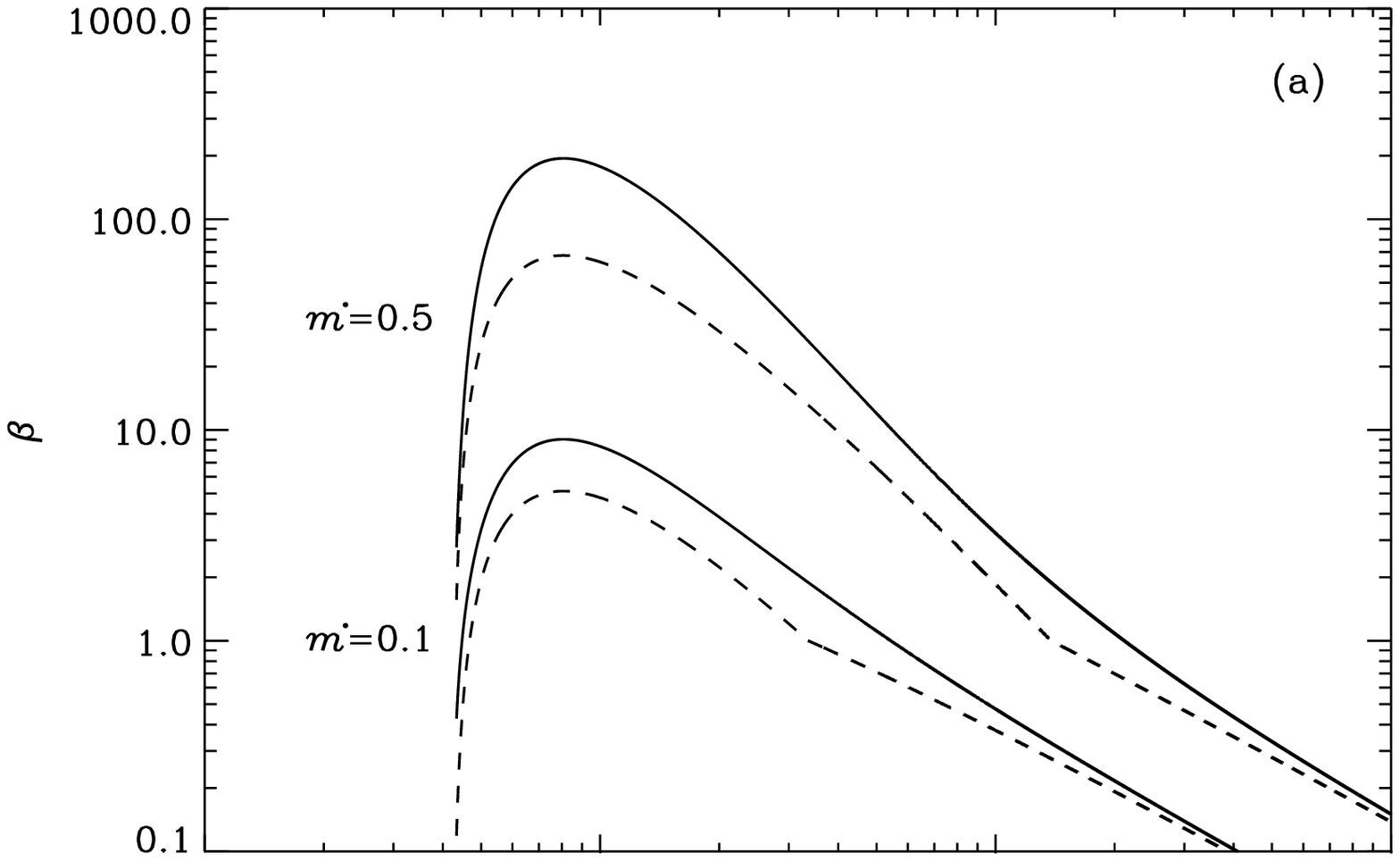}
\includegraphics[width=0.48\linewidth]{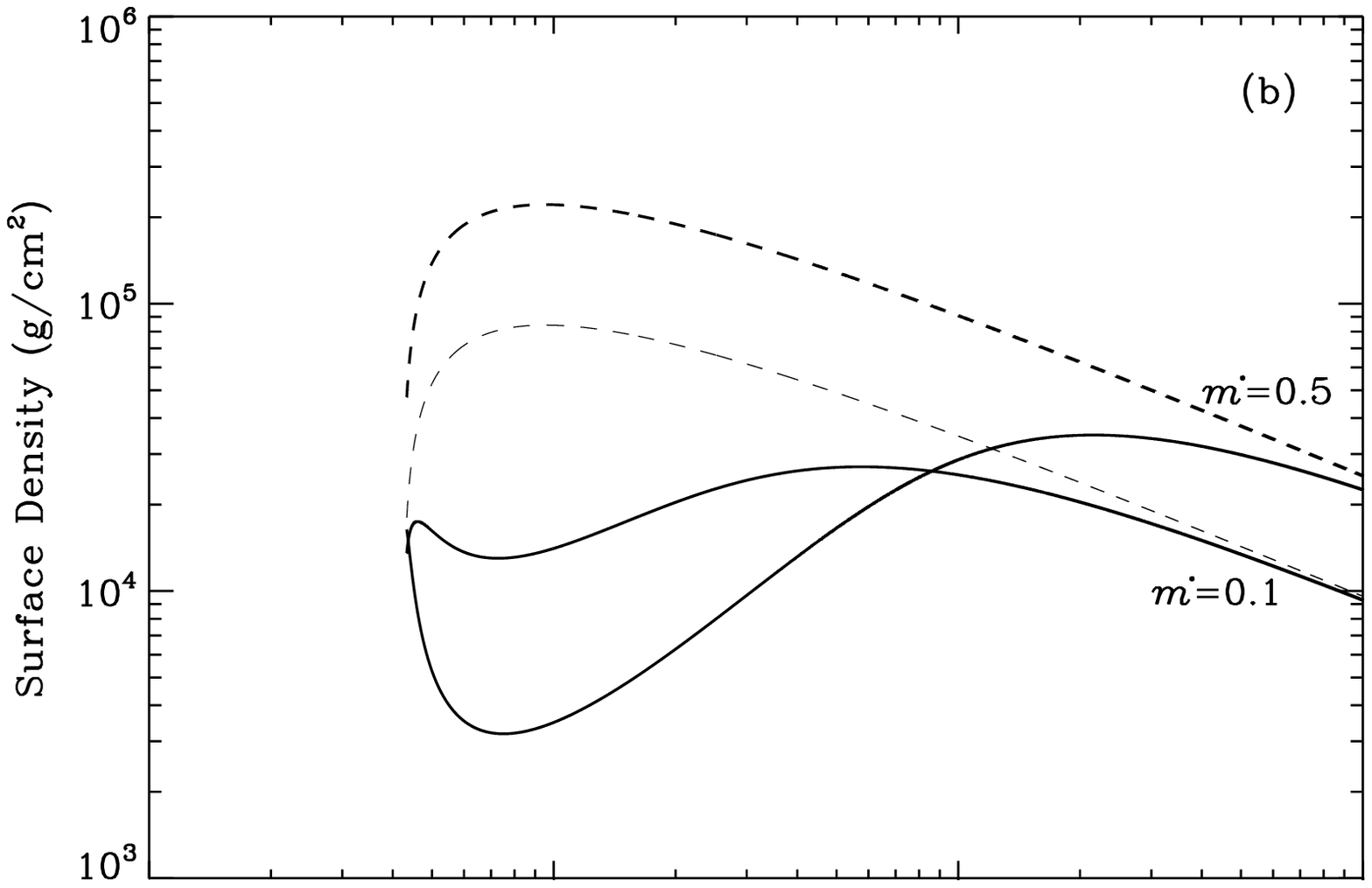} \\
\vspace{-5mm}
\includegraphics[width=0.48\linewidth]{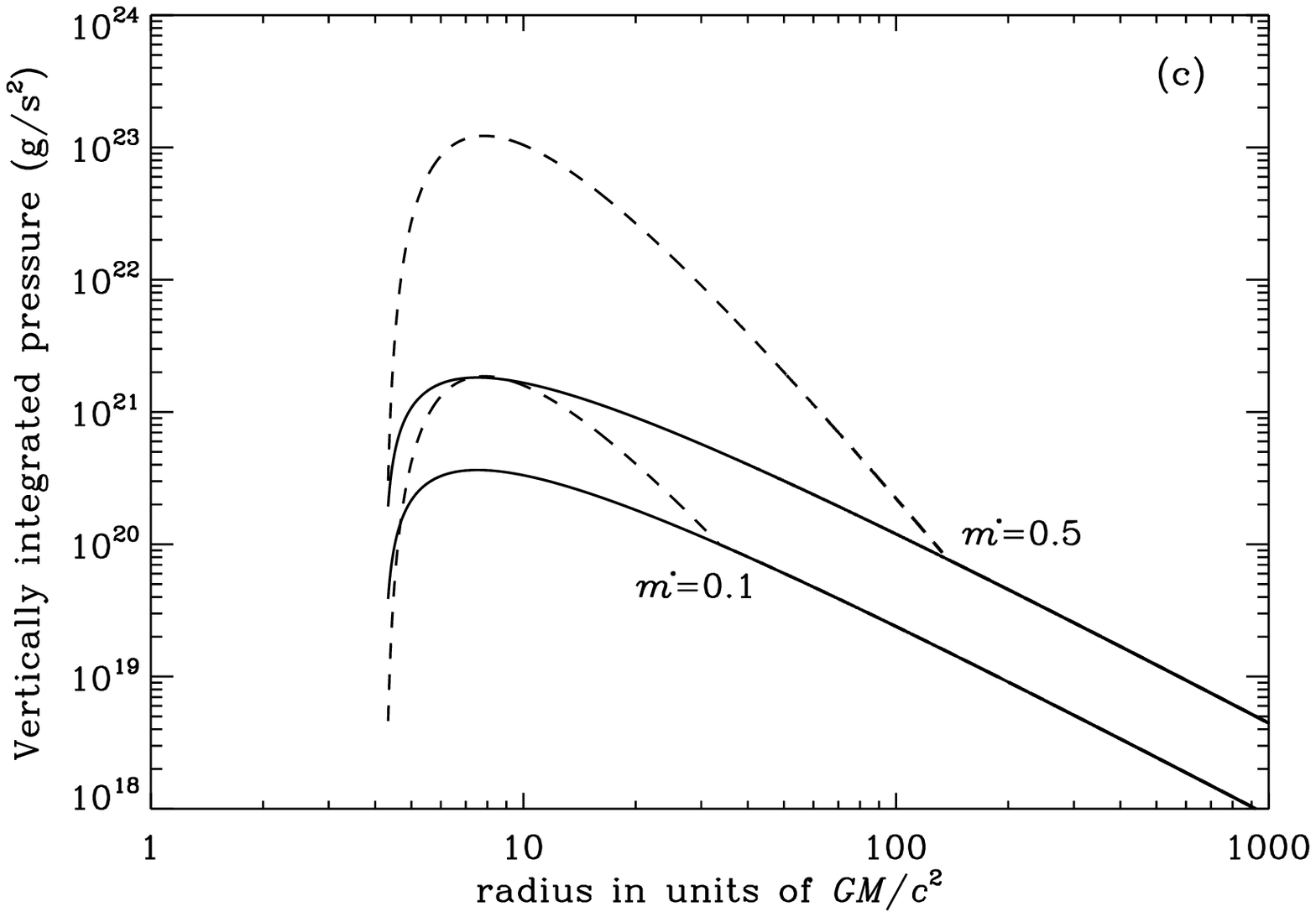}
\includegraphics[width=0.48\linewidth]{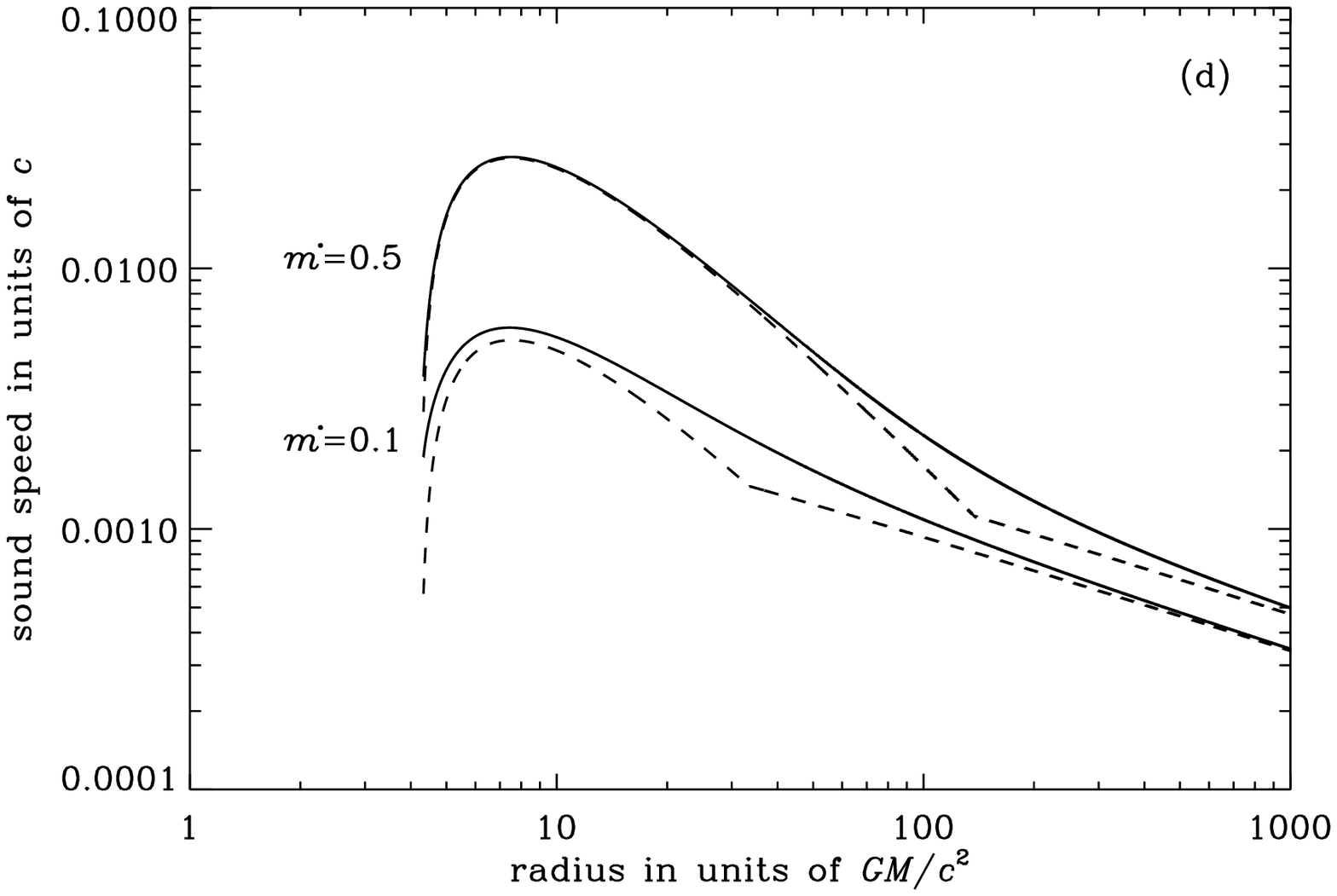}
\end{center}
\caption{Log-log plot of the variation of (a) $\beta=P_\textrm{r}/P_\textrm{g}$, (b) surface density, (c) vertically integrated pressure and (d) sound speed in units of $c$ with radius for $\dot{m}=0.1,\,0.5$, $\alpha=0.1$, $a=0.5$ and $m=10$. The full lines represent the disc model with a combination of gas and radiation pressure when the stress scales with total pressure, while the dashed lines correspond to solutions in regions (a) and (b) when the stress scales with the gas pressure only.}
\label{model}
\end{figure*}

In order to get a smooth transition between the two regions, and to allow for a combination of gas and radiation pressure throughout the disc, we solve equation (\ref{rsa}) numerically. 
The resultant function, $\beta(r)$, is then used in (\ref{rsb}) and (\ref{rsc}), in order to get expressions for $H(r)$ and $\Sigma(r)$. The radial disc structure obtained here is used in the study of the warp and eccentricity propagation, which is described in the following section.

One reason that we take some care over the basic disc model is that it is important to obtain a smooth density profile for the disc in order to calculate the propagation of the warp and eccentricity. Also, since $H^2$ appears in the equations for those quantities, the factor of $1/9$ in equation (\ref{dt}) makes a significant difference.

For comparison, we also include in this section the expressions obtained for region (a), if the stress scales with the gas pressure instead of total pressure ($\mu=\alpha p_g/\Omega$). The semi-thickness of the disc is independent of the viscosity prescription, so that (\ref{ha}) is still valid. The surface density is given by (\ref{sdb}) while $\beta$ and $P$ are given by:

\begin{equation}
P=3.9\times10^{27}\, \dot{m}^{13/5}\alpha^{-4/5}m^{1/5}f^{13/5}r^{-18/5},
\end{equation}
\begin{equation}
\beta =1.3\times10^5\, \dot{m}^{8/5}\alpha^{1/5}m^{1/5}f^{8/5}r^{-21/10}.
\end{equation}

In Fig. \ref{model} we show how $\beta$, the surface density, the vertically integrated pressure and the isothermal sound speed (squared), $c^2_{\mathrm{s}}=p_0/\rho_0=\frac{9}{8}\Omega^2H^2$, vary with radius for a disc with $\alpha=0.1$, $a=0.5$ and $m=10$, for two different values of the accretion rate. The differences in the disc structure when the stress scales with the total pressure or with the gas pressure are more evident in the innermost region of the disc, and for higher accretion rate, i.e., where radiation pressure is more important. The exception is the sound speed, proportional to the thickness of the disc, since $H$ in the radiation-pressure dominated region is independent of the viscosity prescription.

Finally, it should be noted that the model presented in this section is Newtonian. The effect of the black hole appears only in the choice of the inner radius, which is taken to be the marginally stable orbit and is, therefore, dependent on the rotation of the compact object. The relativistic correction factors of \cite{nt73} are not of great importance here since the most important relativistic effects in the study of warp and eccentricity propagation are the apsidal and nodal precession.

\section{Stationary propagation of warp and eccentricity}

In the general case, the warp tilt and the eccentricity are not only functions of space but also of time, and their evolution has been studied by (e.g.) \cite{lubowetal2002} and \cite{ogilvie2001}, respectively. However, the characteristic time-scale for the precession of a global warp or eccentricity is small when compared to the orbital frequency in the binary system, which implies that the frequency of these global deformations is negligible compared to the characteristic frequencies in the inner part of the disc. Therefore, we choose to study the steady shape of a warped or eccentric disc around a black hole.

\subsection{Equations}

In this section we present the equations that are used to describe the stationary wave-like propagation of warp and eccentricity in the disc model presented in the previous section. 

The variation of the tilt of the disc, $W$, with radius can be obtained from \citep{lubowetal2002}

\begin{equation}
\frac{\textrm{d}}{\textrm{d}R}\left[\left(\frac{PR^3\Omega^2}{\Omega^2-\kappa^2+2i\alpha_\textrm{W}\Omega^2}\right)\frac{\textrm{d}W}{\textrm{d}R}\right]=\Sigma R^3(\Omega_z^2-\Omega^2)W.
\label{luboweq}
\end{equation}
This equation describes how propagating bending waves communicate the warp through the disc and is valid for small-amplitude warps; $W$ describes the amplitude and phase of the inclination of the disc. The local wavelength of these bending waves is approximately the one given by the dispersion relation (\ref{disprelation}) for $(\omega,m,n)=(0,1,1)$, if $\alpha_\textrm{W}=0$. The warp propagation is subject to viscous decay, which is described here by a dimensionless viscosity parameter designated $\alpha_\textrm{W}$. In the general case of non-isotropic viscosity, $\alpha_\textrm{W}\neq\alpha$ as the former is related to the $T_{rz}$ and $T_{\phi z}$ components of the stress tensor \textbf{\textsf{T}} while the latter is related to $T_{r\phi}$ \citep[see][and references therein]{lubowetal2002}.

The stationary propagation of a small eccentricity through the disc can be described, in the simplest case, by \citep[e.g.][]{goodchildogilvie2006}

\begin{equation}
\frac{\textrm{d}}{\textrm{d}R}\left[\left(\gamma-\textrm{i}\alpha_\textrm{E}\right)PR^3\frac{\textrm{d}E}{\textrm{d}R}\right]+R^2\frac{\textrm{d}P}{\textrm{d}R}E=\Sigma R^3(\kappa^2-\Omega^2)E,
\label{ecceq}
\end{equation}
where $E$ is a (possibly) complex function representing the amplitude and phase of the eccentricity at a given radius; $\gamma$ is the ratio of specific heats. In a strictly isothermal disc $\gamma=1$ and the global modes described by this equation have a local wavelength which is approximately the one given by the dispersion relation (\ref{disprelation}) for $(\omega,m,n)=(0,1,0)$, if $\alpha_\textrm{E}=0$. The equation describing the eccentricity propagation is based on, and agrees with the local dispersion relation of, a 2D disc; 3D effects are discussed by \cite{ogilvie2001,ogilvie2008}. The parameter $\alpha_\textrm{E}$ in equation (\ref{ecceq}) is essentially a bulk viscosity. Effects of shear viscosity are not included as its direct implementation may lead to growing eccentric waves \citep[viscous overstability,][]{kato1978} \citep[see also][and references therein]{ogilvie2001}. The process of turbulent eccentricity damping is poorly known and the simplest way of describing it is by using a dimensionless bulk viscosity parameter. 

In order to include relativistic effects in the problem, we take $\Omega^{-1}=(GM/c^3) (r^{3/2}+a)$ and use expressions (\ref{relkappa}) and (\ref{relomegaz}) for the radial and vertical epicyclic frequencies, respectively. For the surface density $\Sigma$ and vertically integrated pressure $P$, expressions (\ref{rsc}) and (\ref{rsd}) are used.

\subsection{Numerical method}

We solve equations (\ref{luboweq}) and (\ref{ecceq}) numerically, using a
4th order Runge--Kutta method. In the case of the warp we use the boundary condition that corresponds to zero torque at the inner edge: $\mathrm{d}W/\mathrm{d}R(R_\mathrm{in})=0$. In order for the amplitude of the solution to be fixed, we specify the value $W_0=W(R_\mathrm{in})$ at the inner edge, which corresponds to the (small) inclination of the inner edge of the disc with respect to the equator of the black hole. For the eccentricity we use similar boundary conditions: $\mathrm{d}E/\mathrm{d}R(R_\mathrm{in})=0$ and $E_0=E(R_\mathrm{in})$. To avoid the formal singularity of the equations at the marginally stable orbit, here we take $R_\textrm{in}=R_\textrm{ms}+\delta$, where $\delta\ll R_\textrm{in}$ is an arbitrary value; the solutions for $W(R)$ and $E(R)$ are practically independent of the choice of $\delta$. The solutions are linear and can be renormalized to obtain any desired $W_\textrm{out}$ and $E_\textrm{out}$, the values of disc tilt and eccentricity at the outer radius. This normalisation is particularly meaningful in the case of $W(R)$ since $W_\textrm{out}$ is the (constant) disc tilt at large radius, which can be related to observations.

To solve the warp and eccentricity equations we consider a black hole with 10 solar masses, i.e. $m=10$, and spin $a=0.5$. The ratio of specific heats is given by \citep[e.g.][]{stellar}\footnote{Note that in many references $\beta$ is used to represent the ratio of gas pressure to total pressure while here $\beta=p_\textrm{r}/p_\textrm{g}$.},

\begin{equation}
\gamma=\frac{5+40\beta+32\beta^2}{3+27\beta+24\beta^2}.
\end{equation}
In regions (a) and (b) this expression gives $\gamma\approx4/3$ and $\gamma\approx5/3$, respectively. We regard the dimensionless viscosity parameter $\alpha$ as constant throughout the disc as in \cite{ss73}. Although simulations suggest values of $\alpha$ of the order $10^{-2}$, according to \cite{kingetal2007} observations indicate a typical range of $\alpha\sim0.1-0.4$. Here we choose to fix the viscosity parameter to $0.1$. Parameters $\alpha_\textrm{W}$, $\alpha_\textrm{E}$ and $\dot{m}$ can be varied. To mimic the transition between different black hole states we fix the viscous damping and vary the mass accretion rate. The results and corresponding discussion are presented in the next section.

\section{Results and Discussion}

\subsection{Undamped propagation}

\begin{figure*}
\begin{center}
\includegraphics[width=0.48\linewidth]{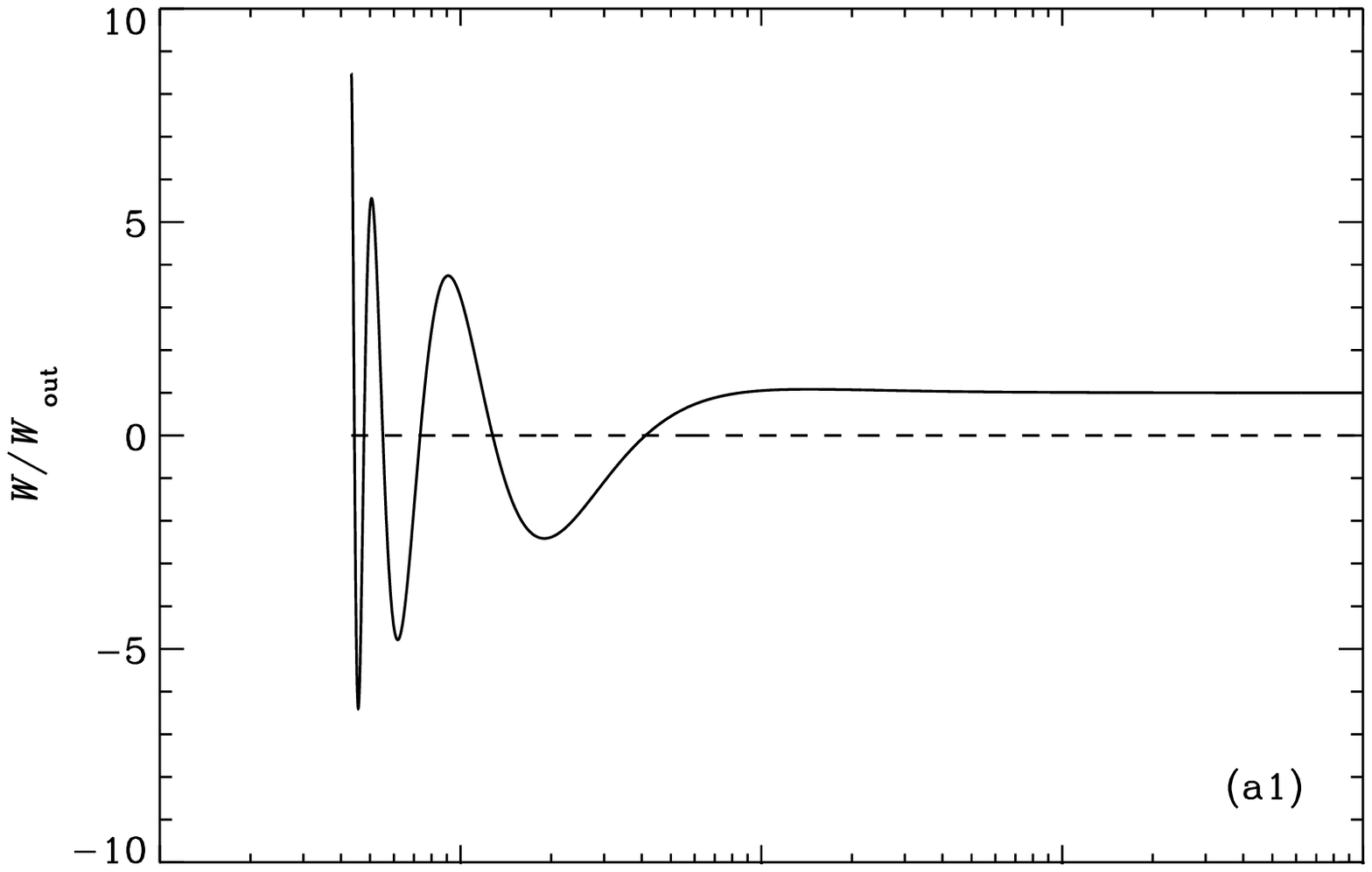}
\includegraphics[width=0.48\linewidth]{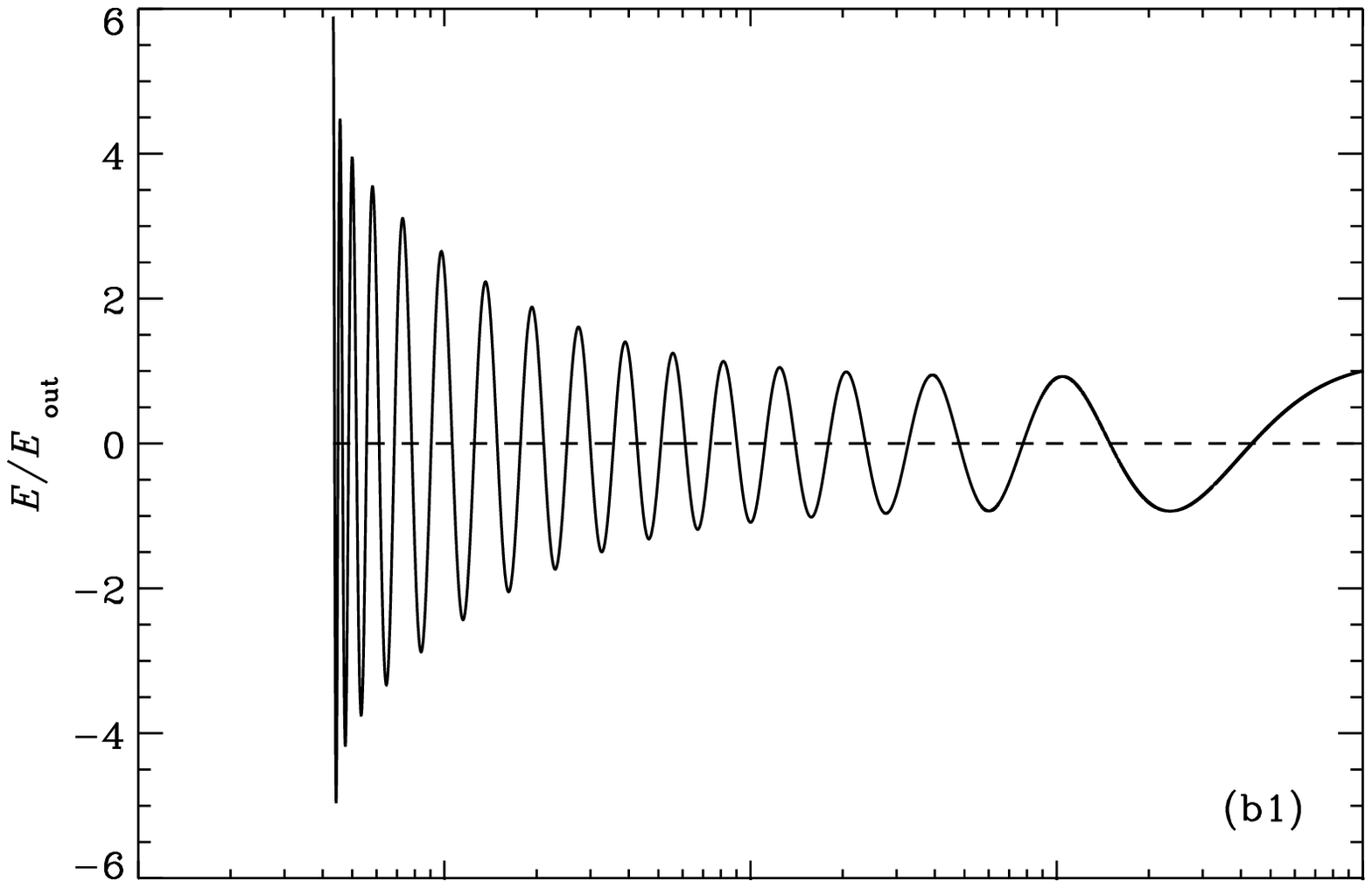} \\
\vspace{-5mm}
\includegraphics[width=0.48\linewidth]{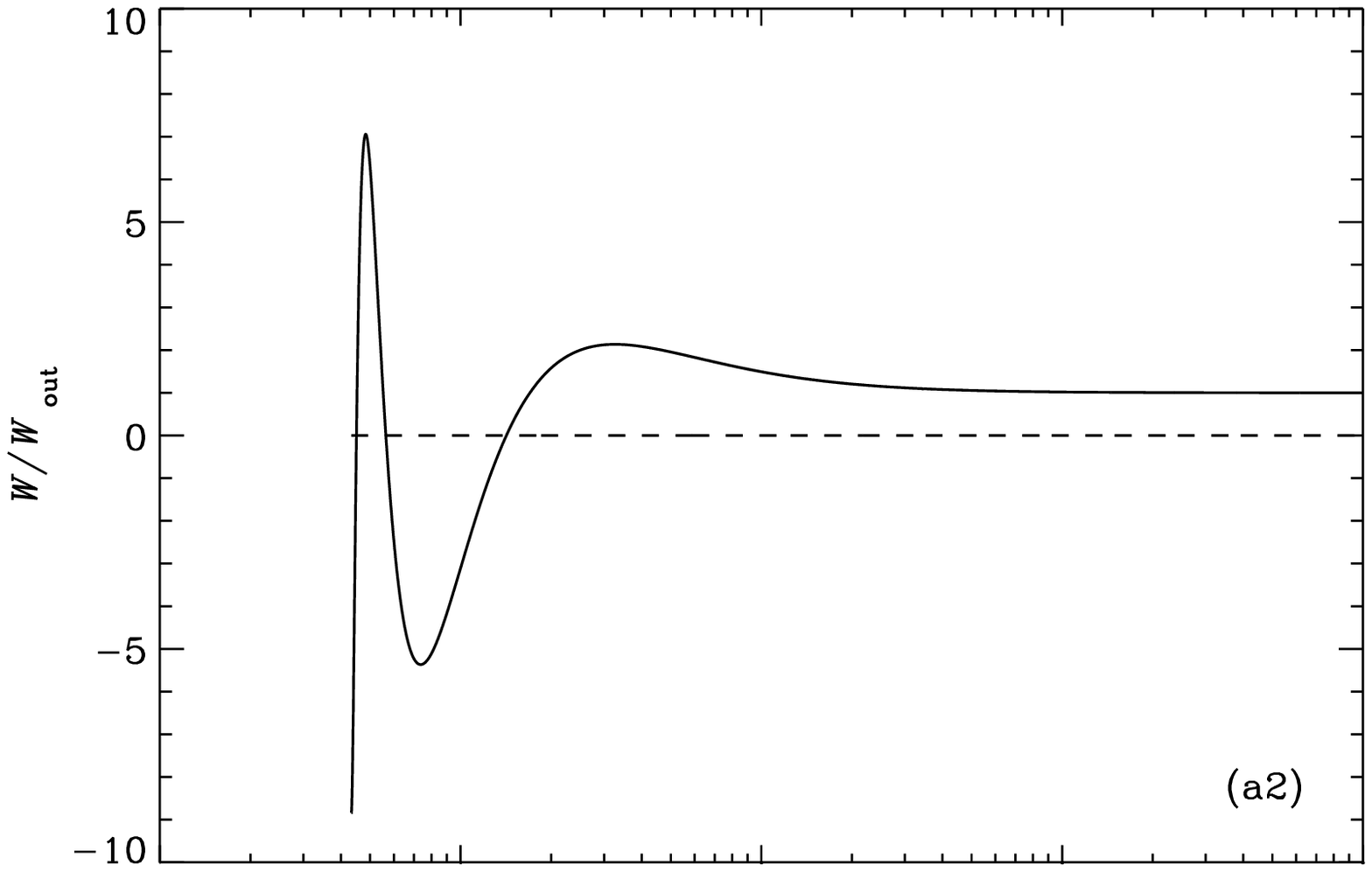}
\includegraphics[width=0.48\linewidth]{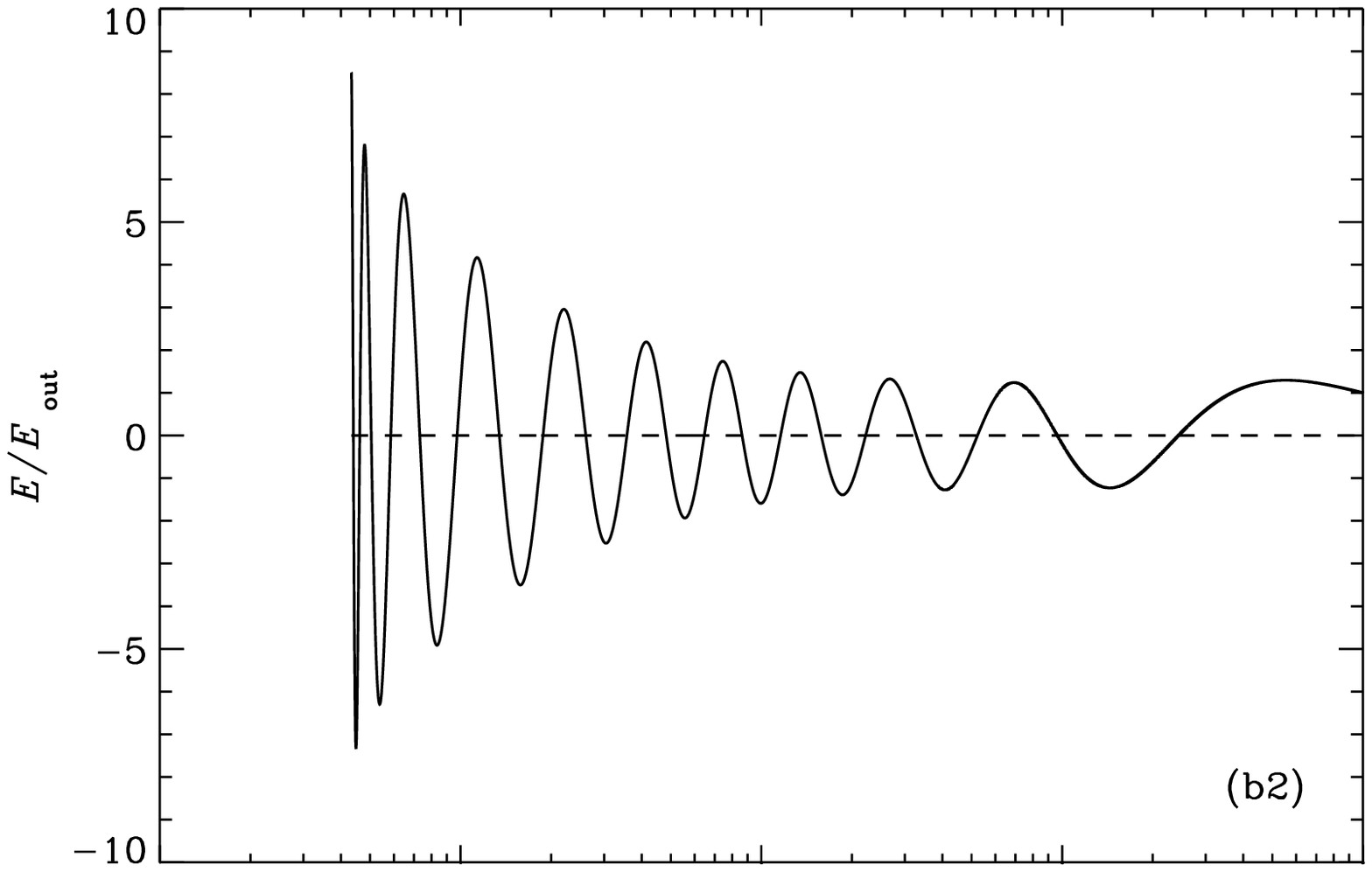} \\
\vspace{-5mm}
\includegraphics[width=0.48\linewidth]{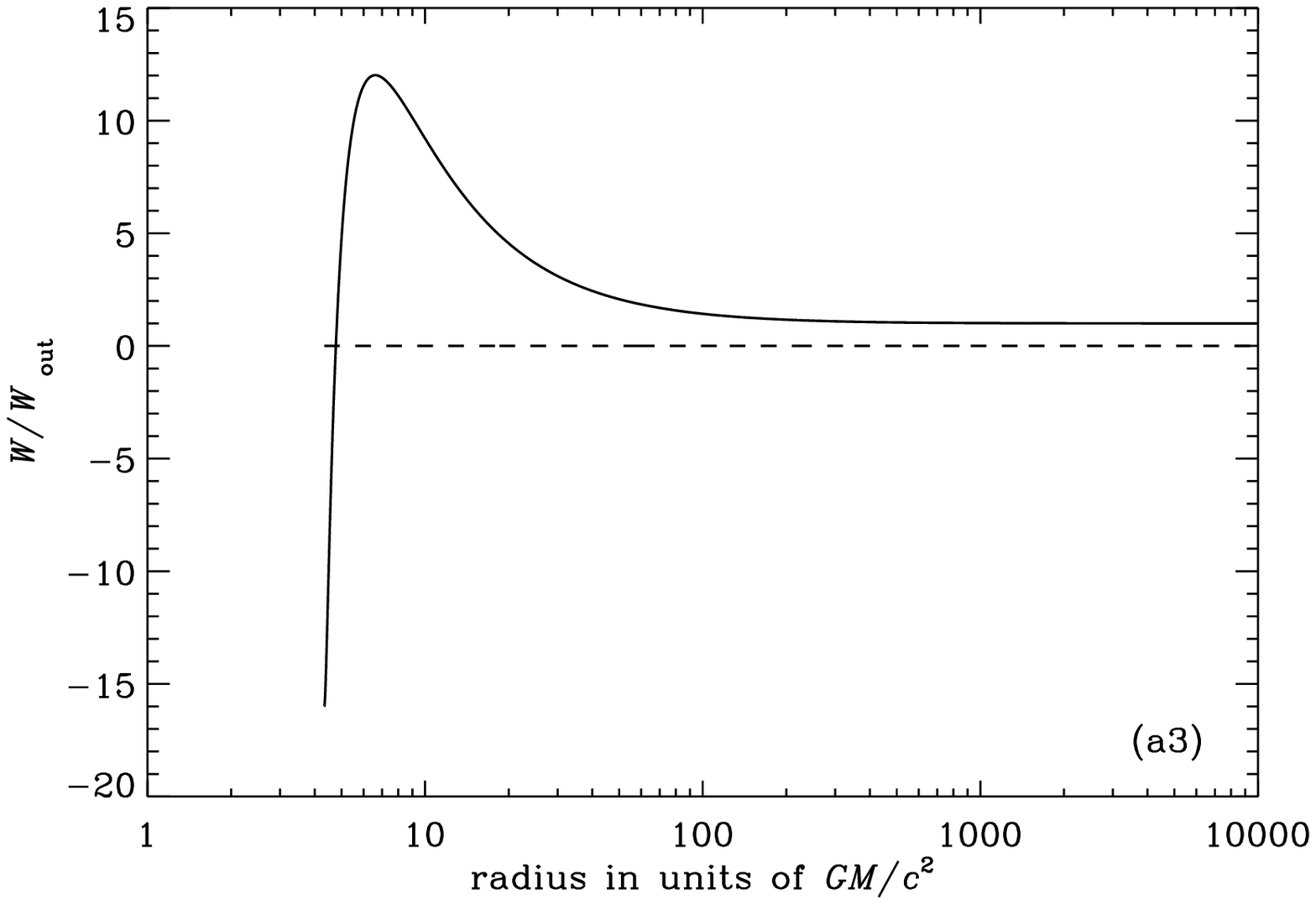}
\includegraphics[width=0.48\linewidth]{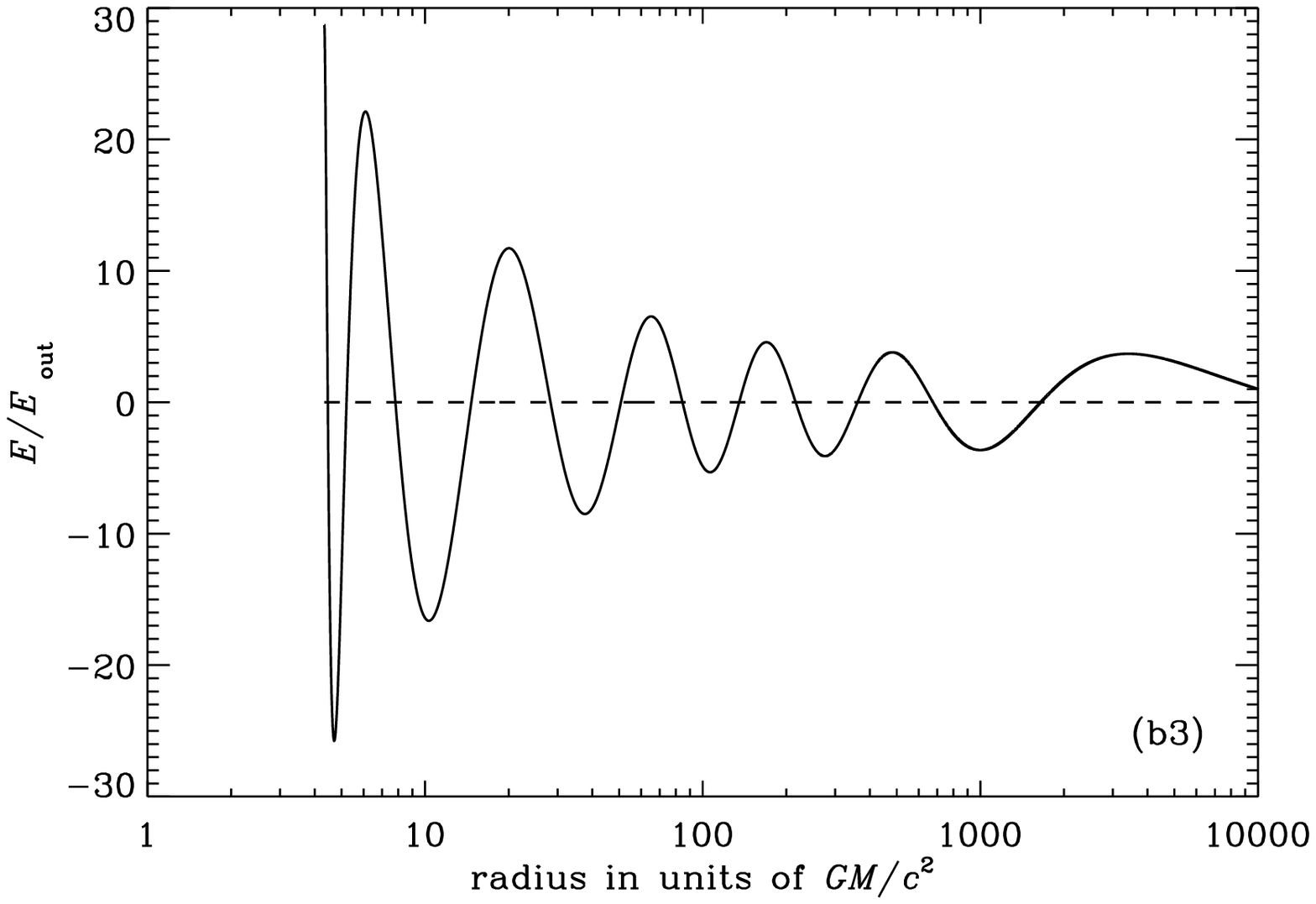}
\end{center}
\caption{Radial variation of (a) the warp tilt and (b) eccentricity normalised to their values at the outer radius for $\alpha_\textrm{W}=0= \alpha_\textrm{E}$ (i.e., no viscous damping) for (1) $\dot{m}=0.2$, (2) $\dot{m}=0.4$, (3) $\dot{m}=0.8$. The full line represents the real part of the disturbance while the imaginary part is represented by the dashed line. A logarithmic scale is used for the x-axis.}
\label{free}
\end{figure*}

\begin{figure*}
\begin{center}
\includegraphics[width=0.48\linewidth]{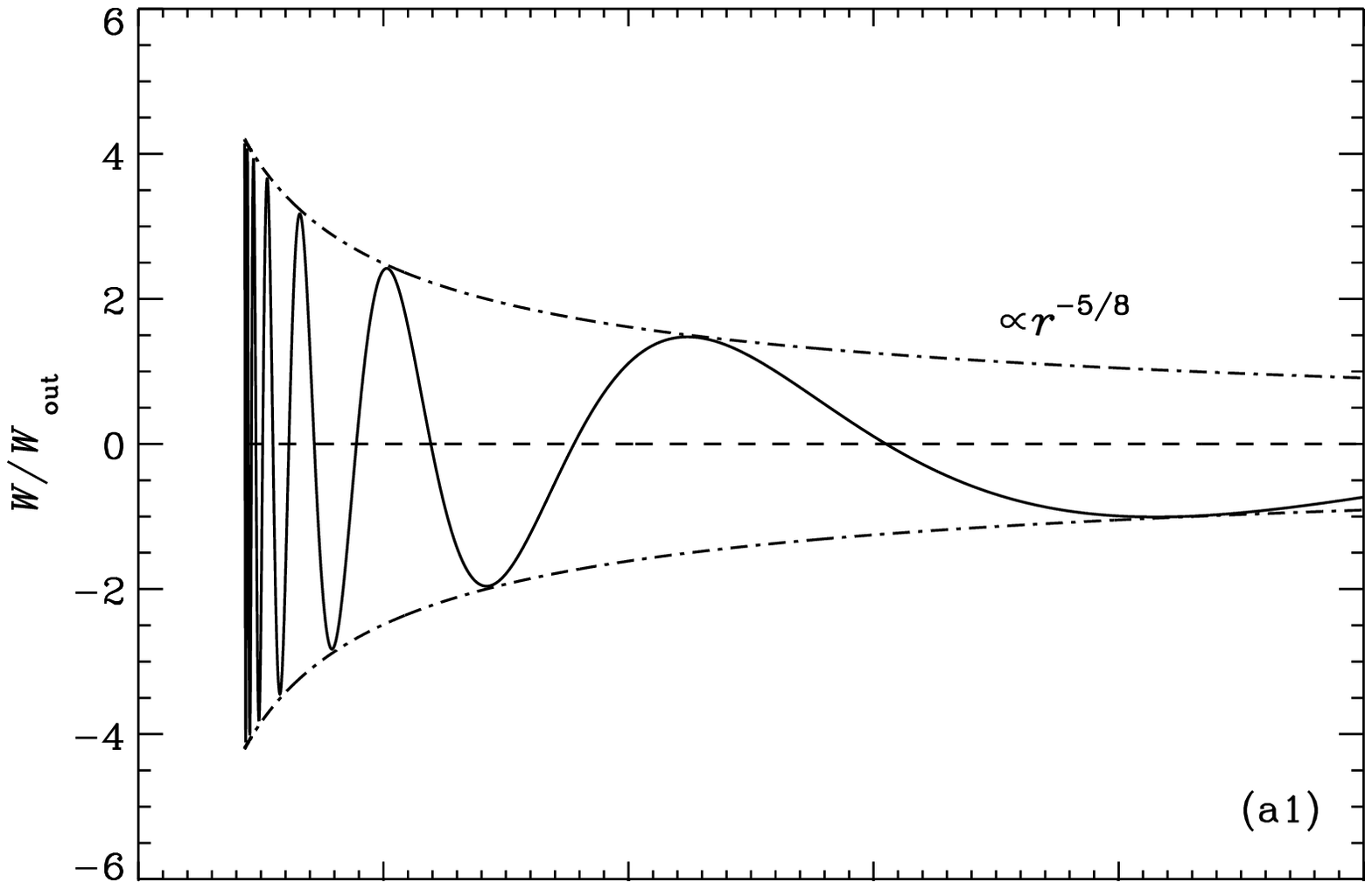}
\includegraphics[width=0.48\linewidth]{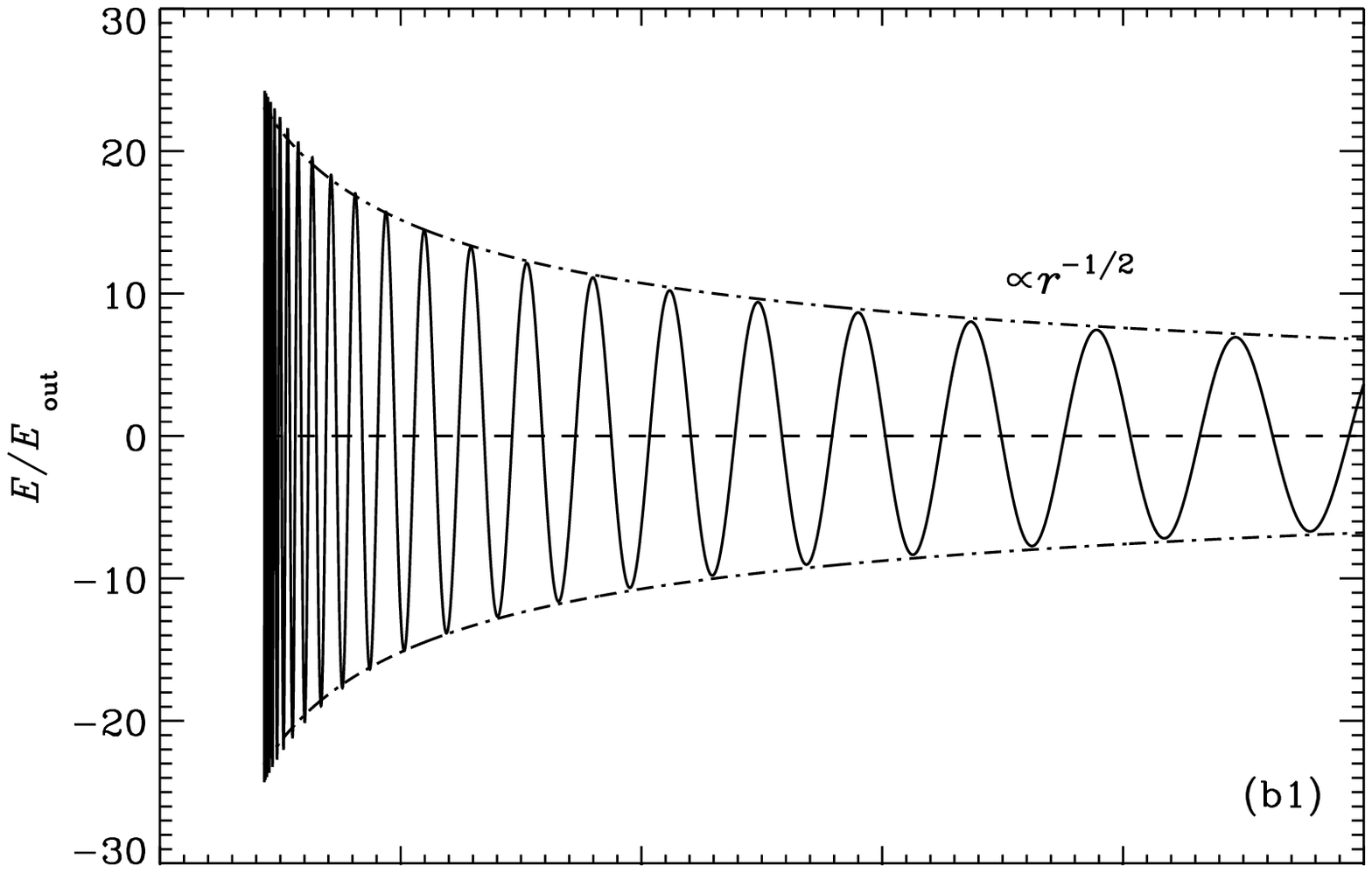} \\
\vspace{-5mm}
\includegraphics[width=0.48\linewidth]{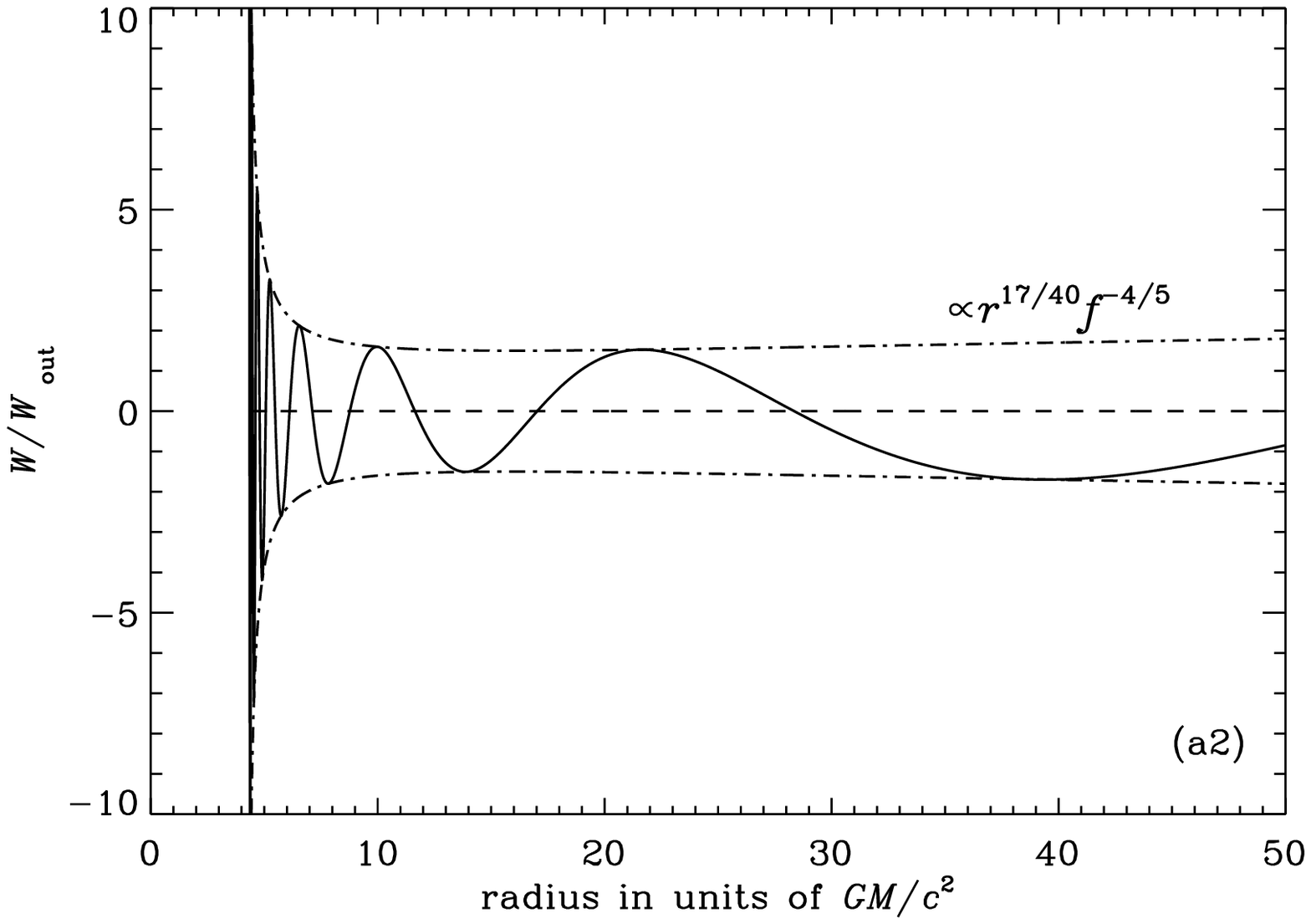}
\includegraphics[width=0.48\linewidth]{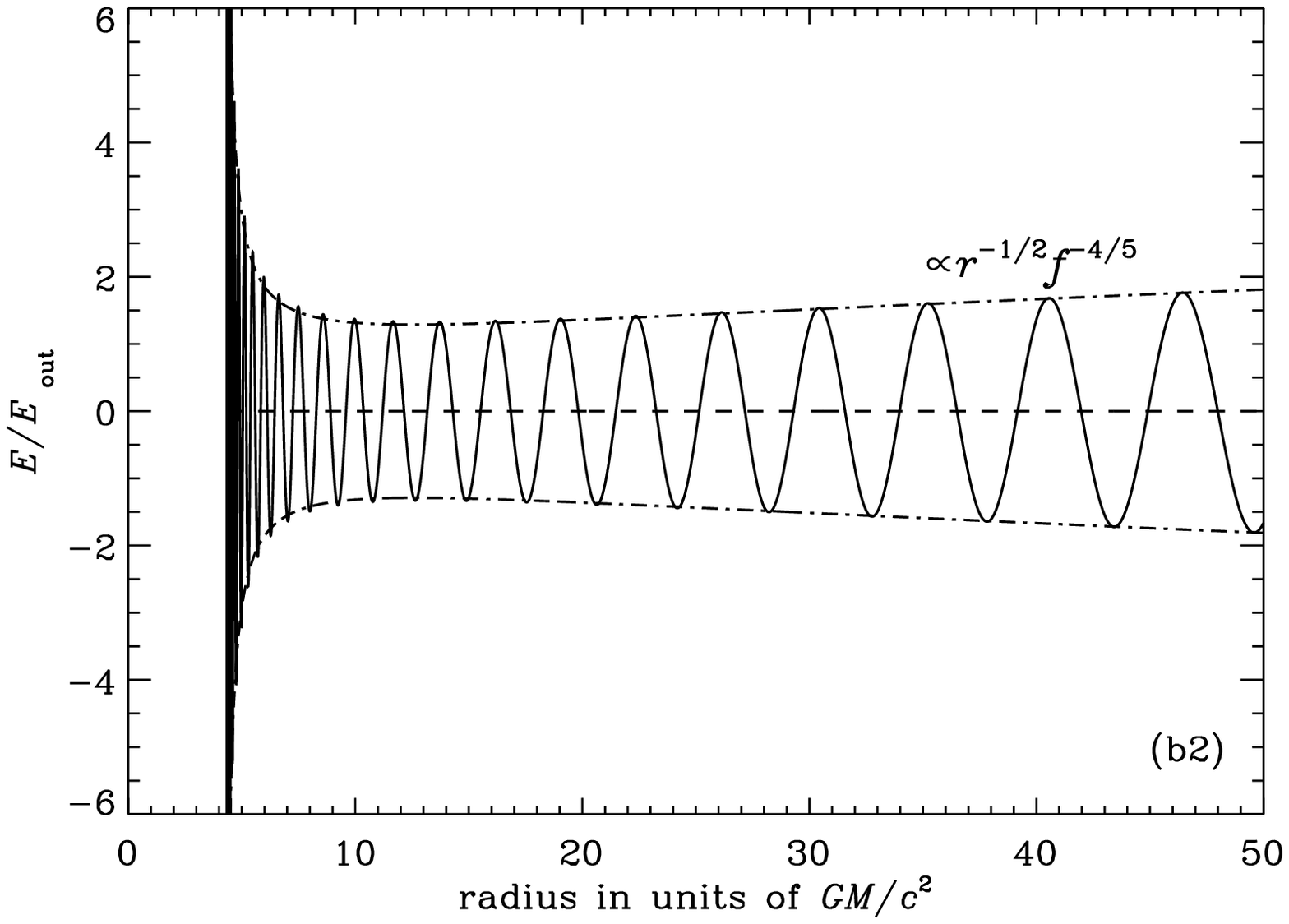}
\end{center}
\caption{Solutions for (a) warp and (b) eccentricity propagation in region (a) ($\beta\gg1$) for $\alpha_\textrm{W}=0= \alpha_\textrm{E}$, $m=10$, $a=0.5$, $\dot{m}=0.1$, when the stress scales with the (1) total pressure and the (2) gas pressure. The full line represents the real part of the disturbance while the imaginary part is represented by the dashed line. The dot-dashed curves show the radial variation of the amplitude of the deformations, as predicted by the WKB analysis.}
\label{wkb}
\end{figure*}

In Fig. \ref{free} we show the variation with radius of the warp tilt and eccentricity for increasing accretion rate when $\alpha_\textrm{W}=0= \alpha_\textrm{E}$. In this case no dissipation is present in the equations for global deformations, so they propagate everywhere with non-negligible amplitude. Since the imaginary part of the solutions is zero, the warp comprises a pure tilt, i.e., the disc is not twisted \citep{lubowetal2002}. For small accretion rate the warp has an oscillatory shape as previously predicted by \cite{ivanovillarionov1997}. The wavelength increases with radius and, at large $R$, $W$ tends to a constant value, the inclination of the outer disc with respect to the equator of the black hole. Note, however, that contrary to the results of \cite{bardeenpetterson1975} the inner disc is not necessarily aligned with the mid-plane of the compact object, as previously shown by \cite{lubowetal2002}. The simulations of \cite{globalsimulations} also found hints of an oscillatory structure characterised by a long-wavelength with no indication of the Bardeen-Petterson effect. Eccentricity also has an oscillatory structure but its wavelength is shorter (yet much longer than the semi-thickness of the disc) and has a slower increase with radius. The shortening of the wavelengths at small radius of both the warp and eccentricity implies a sharp increase of the gradients $\textrm{d}W/\textrm{d}R$ and $\textrm{d}E/\textrm{d}R$ in the inner region of the disc. This is an important feature in terms of wave coupling in this region since the growth rate for inertial modes that are trapped below the maximum of the epicyclic frequency, and interact with global deformations, is proportional to $|\textrm{d}W/\textrm{d}R|^2$ or $|\textrm{d}E/\textrm{d}R|^2$, as found by \cite{ferreiraogilvie2008}.

The basic behaviour of the numerical solutions presented in this section can be explained using WKB theory. In particular, it is possible to predict how the wavelength and amplitude of the several solutions scale with radius. A WKB analysis of the secular theories can be performed by assuming the deformations $W$ and $E$ to be proportional to $\exp(\int\textrm{i}\,b\,\textrm{d}R)$, where $b$ is a function of $R$ to be determined. The $\textrm{d}P/\textrm{d}R$ term in equation (\ref{ecceq}) can be ignored in the WKB analysis since relativistic precession dominates the propagation of eccentricity. Letting $D_\textrm{W}=W(R)$ and $D_\textrm{E}=E(R)$, we can write equations (\ref{luboweq}) and (\ref{ecceq}) in the form

\begin{equation}
\frac{\textrm{d}}{\textrm{d}R}\left(g_i\frac{\textrm{d}D_i}{\textrm{d}R}\right)+h_iD_i=0, \quad i=\textrm{W},\textrm{E}.
\end{equation}
The WKB solution is valid where the radial variation of $g_i$ and $h_i$ is slow and is given by,
\begin{equation}
D_i\approx(g_ih_i)^{-1/4}\exp\left[\pm \int \textrm{i}\sqrt{h_i/g_i}\textrm{d}R\right].
\end{equation}
In the inviscid case ($\alpha_\textrm{W}=0=\alpha_\textrm{E}$), $g_\textrm{W}\approx PR^3 / 6 r^{-1}$, $h_\textrm{W}\approx\Sigma R^3\Omega^24ar^{-3/2}$, $g_\textrm{E}=\gamma PR^3$, and $h_\textrm{E}\approx\Sigma R^3\Omega^26r^{-1}$, keeping only the lowest-order terms in the expressions for $\Omega^2-\kappa^2$ and $\Omega^2-\Omega_z^2$. Remembering that $P=\Omega^2H^2\Sigma$ (from \ref{he2}), we can write 

\begin{equation}
W\propto r^{1/8}(\Sigma H)^{-1/2}\exp\left[\pm \int \textrm{i}\sqrt{\frac{24a}{r^{5/2}H^2}}\textrm{d}R\right],
\label{wwkb}
\end{equation}
\begin{equation}
E\propto r^{1/4}(\Sigma H)^{-1/2}\exp\left[\pm \int \textrm{i}\sqrt{\frac{6}{\gamma r H^2}}\textrm{d}R\right],
\label{ewkb}
\end{equation}
where we have ignored the variation of $\gamma$ with radius. If the stress scales with the total pressure, these expressions indicate that the amplitude of the warp should be proportional to $r^{-5/8}$ in region (a) and to $r^{-1/10}f^{-2/5}$ in region (b); the amplitude of the eccentricity is proportional to $r^{-1/2}$ in the radiation-pressure dominated regime and to $r^{1/40}f^{-2/5}$ when gas pressure dominates. If the stress scales with the gas pressure only, the amplitudes in region (a) are proportional to $r^{17/40}f^{-4/5}$ and $r^{11/20}f^{-4/5}$ in the case of the warp and eccentricity, respectively. The results for region (a) can be confirmed in Fig. \ref{wkb}. In region (b) $r\gg r_{\textrm{in}}$, $f\approx 1$, so that the radial variation of both warp and eccentricity amplitudes is slow, as seen in Fig. \ref{free}. The eccentricity does not tend to a constant at large radius because its amplitude is slowly increasing when $r$ increases. The warp amplitude decreases with $r$ but does not tend to zero as indicated by the WKB scaling. The reason for this apparent contradiction is in the failure of the WKB approximation at large radii where the wavelength tends to infinity.

The radial variation of the wavelength of the deformations can also be understood in light of the WKB theory. From equations (\ref{wwkb},\ref{ewkb}), we see that the wavelengths are given by $\lambda_\textrm{W}\approx H \pi r^{5/4}/\sqrt{6 a}$, and $\lambda_\textrm{E}\approx 2H\pi r^{1/2}\sqrt{\gamma/6}$, for the warp and eccentricity respectively. These dependencies are in agreement with the results shown in Fig. \ref{free}, where it is clear that the wavelengths of the deformations increase with radius and with the disc thickness (or equivalently with the accretion rate, cf. (\ref{ha},\ref{hb})). The increase in $\lambda$ when $\dot{m}$ increases is evident: in particular, for $\dot{m}=0.8$ the wavelength of the warp becomes so large that it practically loses its oscillatory character, but still maintains the sharp increase in $\textrm{d}W/\textrm{d}R$. Even in this case, the inner disc is not aligned with the equator of the black-hole as in the \cite{bardeenpetterson1975} picture. Unfortunately the behaviour of the solutions very close to the marginally stable orbit cannot be trusted in detail because the singularity of the disc model there ought to be resolved by the transition to a supersonic inflow.

\subsection{Damped propagation}

\begin{figure*}
\begin{center}
\includegraphics[width=0.48\linewidth]{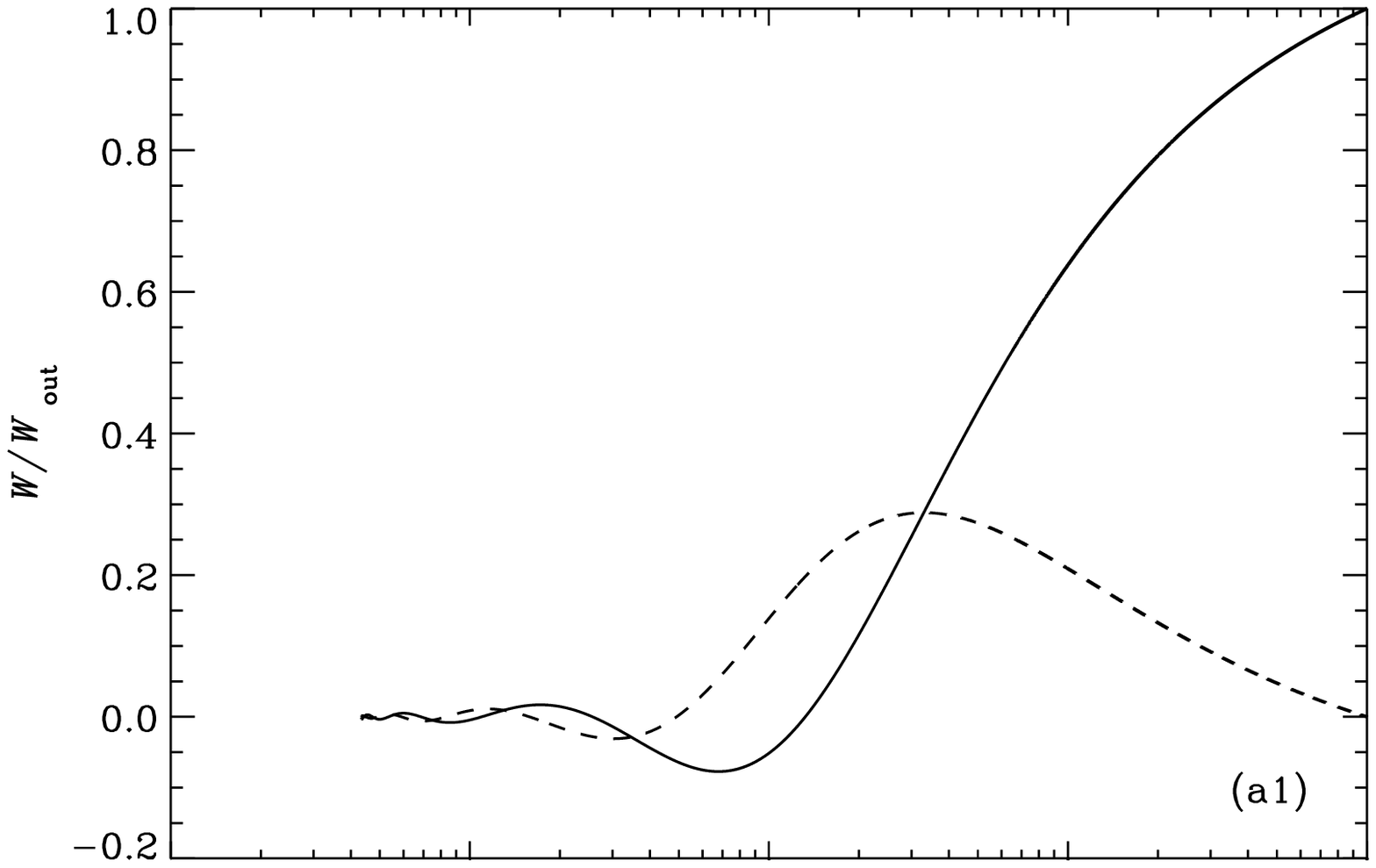}
\includegraphics[width=0.48\linewidth]{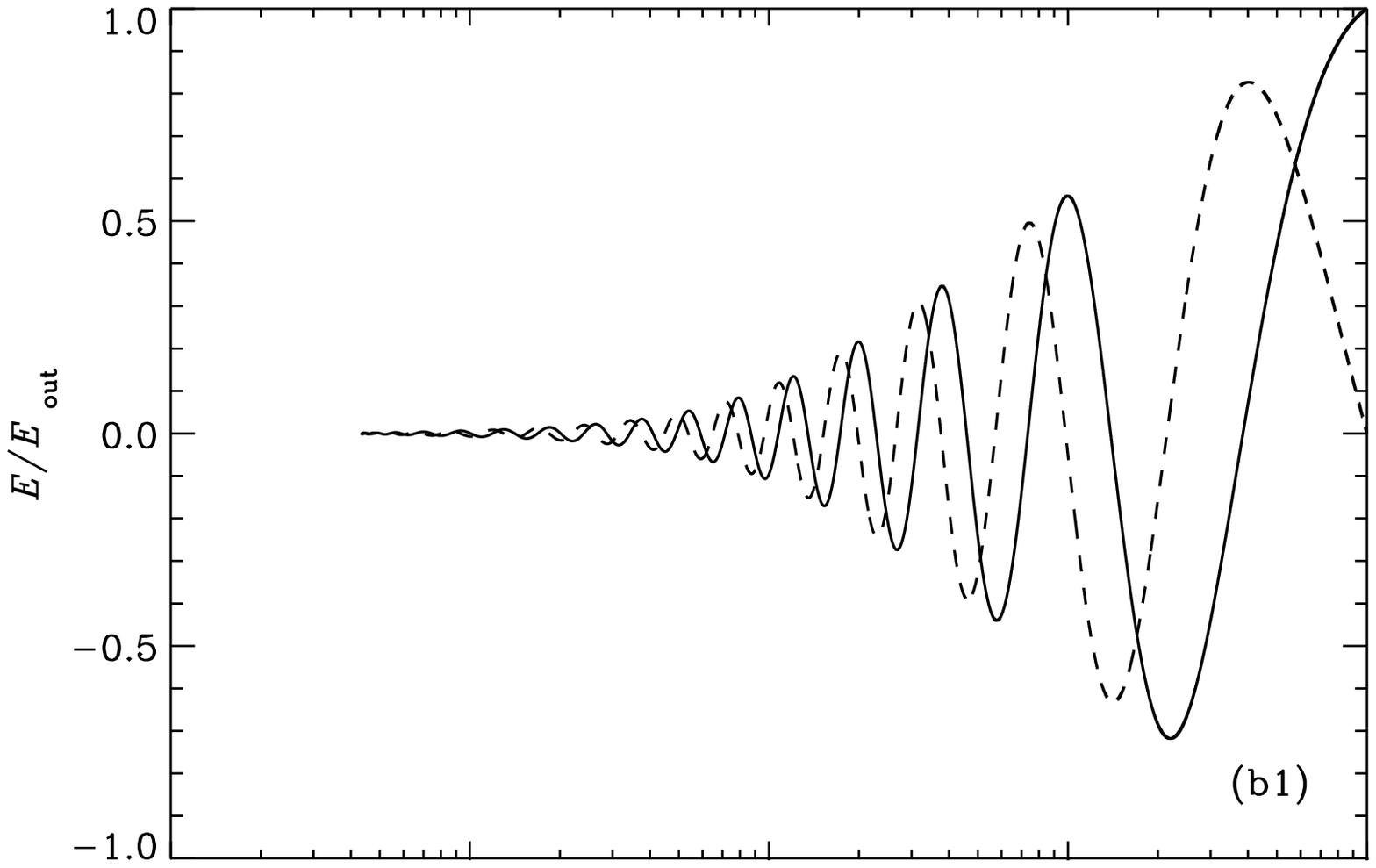} \\
\vspace{-5mm}
\includegraphics[width=0.48\linewidth]{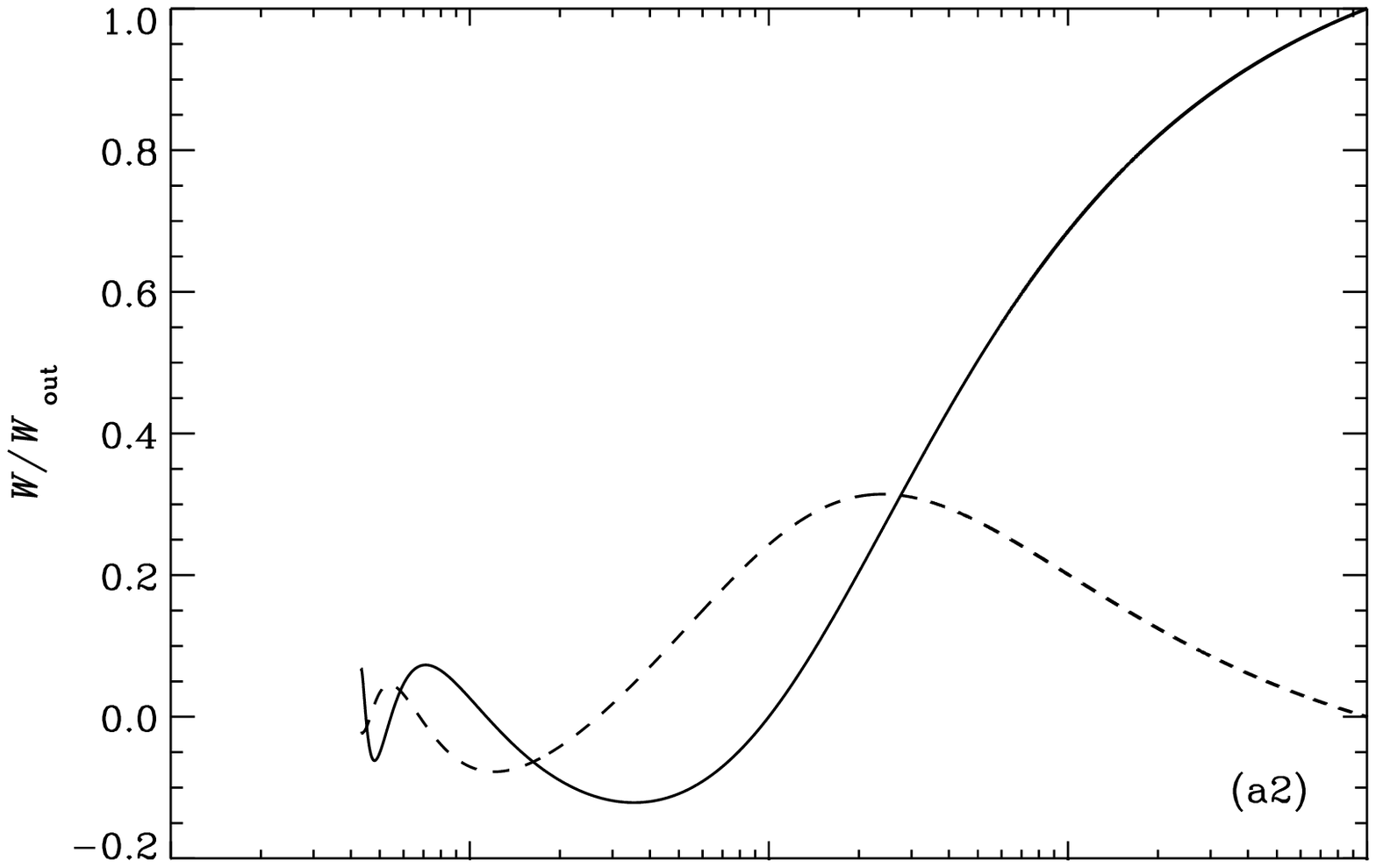}
\includegraphics[width=0.48\linewidth]{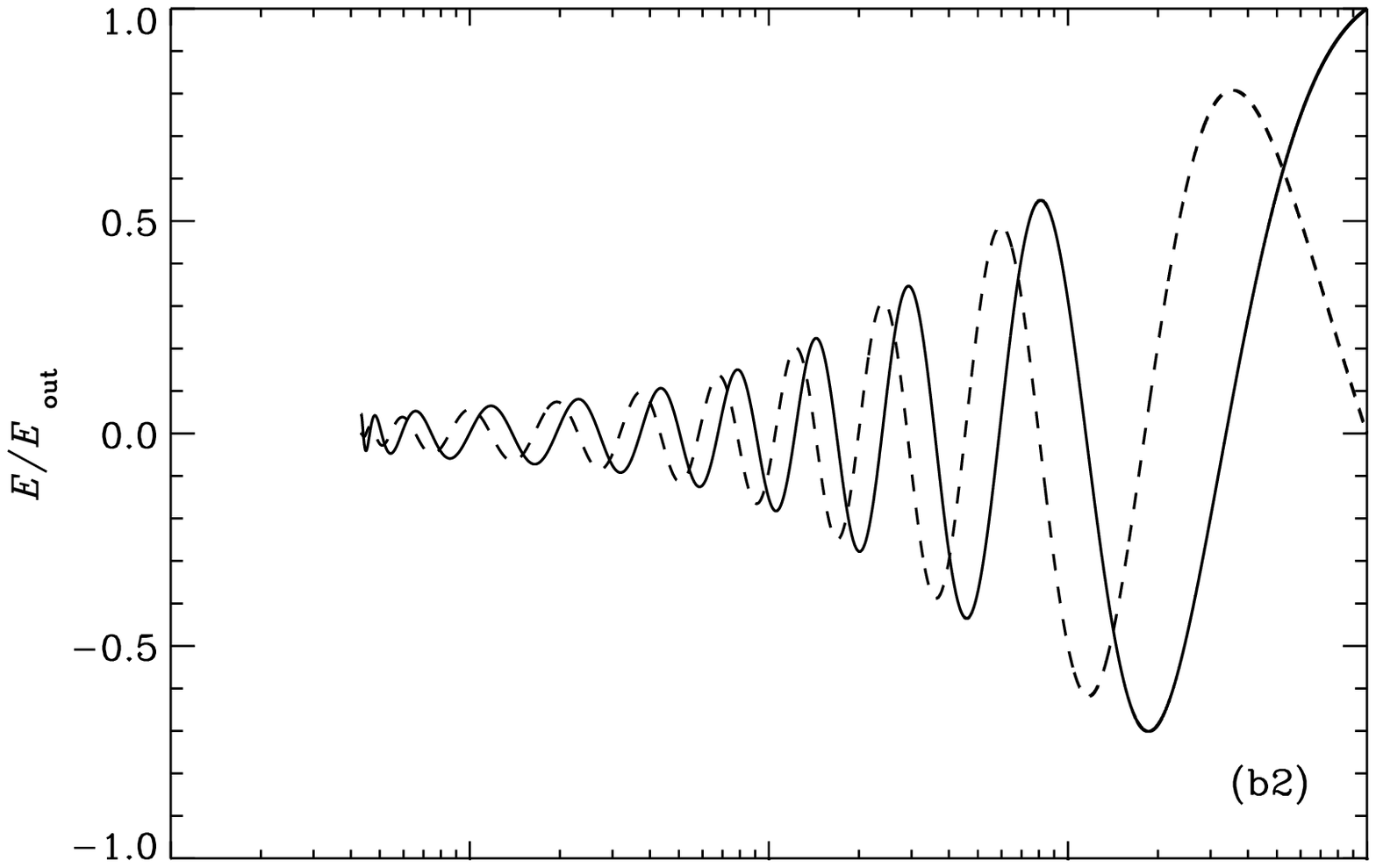} \\
\vspace{-5mm}
\includegraphics[width=0.48\linewidth]{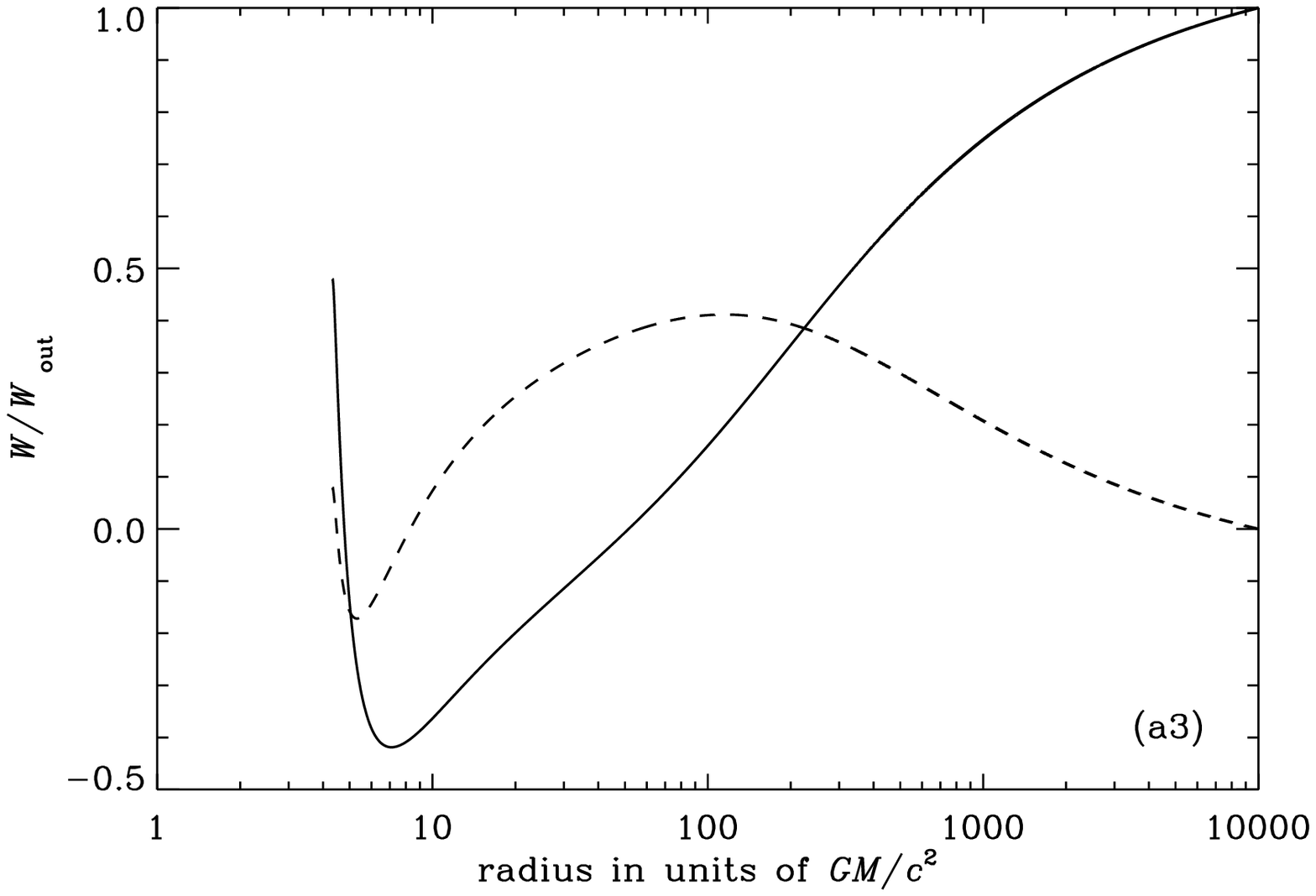}
\includegraphics[width=0.48\linewidth]{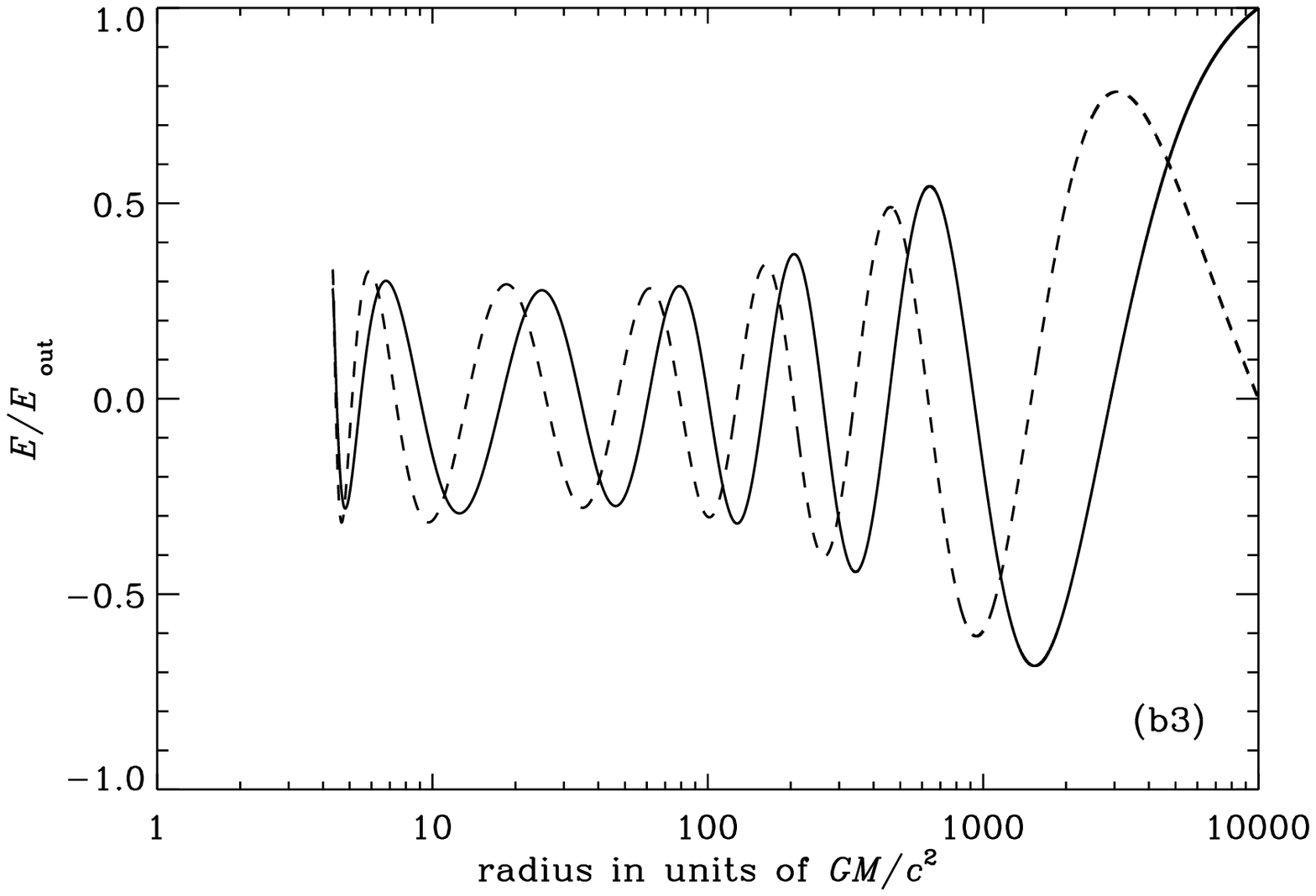}
\end{center}
\caption{Radial variation of (a) the warp tilt and (b) eccentricity normalised to their values at the outer radius for $\alpha_\textrm{W}=0.15$ and $\alpha_\textrm{E}=0.25$ for (1) $\dot{m}=0.2$, (2) $\dot{m}=0.4$, (3) $\dot{m}=0.8$. The full line represents the real part of the disturbance while the imaginary part is represented by the dashed line. A logarithmic scale is used for the x-axis.}
\label{damp}
\end{figure*}

In Fig. \ref{damp} we show the variation with radius of the warp tilt and eccentricity in the case where a significant attenuation is present. The values of $\alpha_\textrm{W}=0.15$ and $\alpha_\textrm{E}=0.25$ are both larger than the viscosity value $\alpha$ used to model the disc, and are chosen so that the warp and eccentricity have negligible amplitude at the inner region of the disc, for small accretion rate ($\dot{m}=0.2$). For dissipation values smaller than these ones, the global deformations propagate everywhere with non-negligible amplitude even for small accretion rate, although with larger amplitude in the inner region for larger $\dot{m}$. 

An important difference between the results presented here and the ones in Fig. \ref{free} is the existence of a non-zero imaginary part in the solutions. This shows that the reflection of the global deformations from the stress-free boundary is no longer perfect since they are affected by viscous attenuation and, in consequence, their amplitude is reduced as they propagate to smaller radii; a wave with inward group velocity, as opposed to a standing wave, is set up. 

It is evident from Fig. \ref{damp} that when the accretion rate increases the global deformations propagate to the inner region of the disc more easily, i.e., the amplitude of warp and eccentricity in the inner region increases with $\dot{m}$. This result can be explained in light of the WKB analysis introduced in 4.1. If a small viscosity is present ($\alpha_\textrm{W}, \, \alpha_\textrm{E}\neq 0$), the WKB solution for the warp and eccentricity includes an attenuation factor due to the presence of a imaginary wavenumber $k_\textrm{i}$,

\begin{equation}
k_\textrm{i}^\textrm{Warp}\approx \pm\frac{\alpha_\textrm{W}}{3H}\sqrt{\frac{6a}{r^{1/2}}}, \qquad k_\textrm{i}^\textrm{Ecc}\approx \pm\frac{\alpha_\textrm{E}}{2\gamma H}\sqrt{\frac{6}{\gamma r}}.
\end{equation}
If $H/r$ varies slowly with radius, the logarithm of the attenuation factor is

\begin{equation}
\int^{\infty}_{R_\textrm{in}} |k_\textrm{i}| \textrm{d}R\sim\frac{\alpha_i}{H/r} \quad (\alpha_i=\alpha_\textrm{W},\,\alpha_\textrm{E}),
\end{equation}
indicating that, for fixed viscosity, the attenuation is smaller if the thickness of the disc (or equivalently $\dot{m}$) is larger.

The results presented in this section indicate that global deformations can reach the inner region of accretion discs under a wide variety of conditions. In particular, even when subject to large viscous attenuation, both warp and eccentricity can reach the inner disc provided the black hole is accreting mass at a large enough rate.

Some caveats accompany the solutions obtained for $W(R)$ and $E(R)$. The equations used assume that these quantities are small enough so that non-linear effects can be neglected. They also assume that the quantities $|1-\Omega_z^2/\Omega^2|$ and $|1-\Omega_z^2/\Omega^2|$ are smaller than, or of the order of, $H/r$ which is not true of the inner region of relativistic discs. Another important caveat relates to the effects of viscosity in the propagation of global deformations. Here they are parametrised using viscous coefficients $\alpha_\textrm{W}$ and $\alpha_\textrm{E}$; this is the simplest way of describing the poorly understood process of turbulent damping in accretion discs. In the particular case of the eccentricity equations only a bulk viscosity is present so that the complications of viscous overstability can be avoided. However, the latter may be present resulting in growth (as opposed to decay) of eccentricity at small radii.

\subsection{Relation to global modes}

The above results are for strictly stationary deformations such as may be induced by a misalignment between the rotational axis of the binary orbit and the spin axis of the black hole. We have also computed slowly precessing global eccentric modes in binary stars by a method similar to \cite{goodchildogilvie2006} but including the relativistic expressions for the characteristic frequencies. When relativistic precession is included, the global modes have almost the same precession and decay rates as in a Newtonian model, while the solution in the inner part of the disc resembles the stationary solutions described above.

\section{Conclusion}

In this paper we studied the propagation of warp and eccentricity in discs around black holes to determine the conditions under which these disturbances can propagate to the inner regions of accretion discs. High-frequency QPOs have previously been identified with inertial oscillations trapped in the inner region of discs, and are detected mainly when black holes are in the very high state where accretion rate is maximum. We find the accretion rate to have a vital role in the damped propagation of global deformations. Our results suggest that the activation of the inner region, and consequent excitation of trapped oscillations by these disturbances, may be possible only when the accretion rate is close to its Eddington value, i.e., when the black hole is in the very high state.

When the propagation of global disturbances (found to have an oscillatory structure in the radial direction) is not affect by viscous damping, the increase in mass accretion rate gives rise to a lengthening in their wavelength, in agreement with the WKB analysis of the warp and eccentricity equations. The most interesting results of our calculations are obtained when the more realistic situation of damped propagation of global disturbances is considered. In this case, the increase of accretion rate facilitates the propagation of warp and eccentricity, i.e., their amplitudes in the inner region increase with $\dot{m}$. In particular, when the accretion rate is only a small fraction of Eddington and the viscous damping is strong enough to completely suppress the propagation of global deformations, an increase in $\dot{m}$ to the values expected in the very high state results in their amplitude in the inner region being increased to a non-negligible value. 



A limitation of the disc model used in our calculations and introduced in section 2 is the fact that it does not include relativistic effects. For example, the law for angular momentum conservation in relativistic discs \citep{nt73} is slightly different from the one used here (cf \ref{cam}). However, relativistic corrections are small and their inclusion is not expected to significantly affect the final results. Also, since the terms on the right-hand side of equations (\ref{luboweq},\ref{ecceq}) dominate the warp and eccentricity propagation, it is more important to introduce relativistic effects in these equations by using expressions (\ref{relomega})--(\ref{relomegaz}) for the characteristic frequencies, as they correctly describe apsidal and nodal relativistic precession rates, related to $\Omega_z^2-\Omega^2$ and $\kappa^2-\Omega^2$.

\section*{Acknowledgments}

We thank Omer Blaes and Chris Fragile for helpful discussions. The work of BTF was supported by FCT (Portugal) through grant no. SFRH/BD/22251/2005.

\label{lastpage}

\end{document}